\documentclass[superscriptaddress,prd,twocolumn,aps,amssymb,nofootinbib]{revtex4}

\usepackage{graphicx}
\usepackage{amssymb}

\newcommand{\Madm}{M_{\rm ADM}}
\newcommand{\MK}{M_{\rm K}}
\newcommand{\beq}{\begin{equation}} 
\newcommand{\eeq}{\end{equation}} 
\newcommand{\beqn}{\begin{eqnarray}} 
\newcommand{\eeqn}{\end{eqnarray}} 
\newcommand{\pa}{\partial}
\newcommand{\na}{\nabla}
\newcommand{\gab}{g^\alpha\!_\beta}

\newcommand{\gabd}{g_{\alpha\beta}}

\newcommand{\gmabd}{\gamma_{ab}}

\newcommand{\fabd}{f_{ab}}

\newcommand{\tbeta}{\tilde{\beta}}

\newcommand{\Aabd}{A_{ab}}

\newcommand{\albe}{{\alpha\beta}}

\newcommand{\Tabd}{T_{\alpha\beta}}

\newcommand{\Tab}{T^\alpha{}\!_\beta}
\newcommand{\Tba}{T_\alpha{}^\beta}

\newcommand{\Gabd}{G_{\alpha\beta}}

\newcommand{\Rab}{R^\alpha{}\!_\beta}

\newcommand{\zD}{{\raise1.0ex\hbox{${}^{\ \circ}$}}\!\!\!\!\!D}
\newcommand{\alone}{{\raise0.5ex\hbox{${}^{\ 1}$}}\!\!\!\!\alpha}
\newcommand{\Od}{{O}}

\newcommand{\dl}{\delta}

\newcommand{\Dl}{\Delta}

\newcommand{\Lie}{\mbox{\pounds}}

\newcommand{\compa}{M/R}

\newcommand{\nalam}{\mathrel{\raise0.9ex\hbox{$^\lambda$}\mkern-14mu
\lower0.0ex\hbox{$\nabla$}}}

\newcommand{\gmaa}{\gamma_a\!{}^\alpha}

\newcommand{\tomega}{\tilde\omega}
\newcommand{\tphi}{\tilde\phi}

\newcommand{\rhoH}{\rho_{\rm H}}

\newcommand{\tA}{\tilde A}

\newcommand{\Kabd}{K_{ab}}

\newcommand{\zeroD}{{\raise1.0ex\hbox{${}^{\ \circ}$}}\!\!\!\!\!D}
\newcommand{\zLap}{{\raise1.0ex\hbox{${}^{\ \circ}$}}\!\!\!\!\Delta}
\newcommand{\zna}{{\raise1.0ex\hbox{${}^{\ \circ}$}}\!\!\!\!\!\nabla}
\newcommand{\zS}{{\raise1.0ex\hbox{${}^{\ \circ}$}}\!\!\!\!\!S}

\newcommand{\bR}{{\bar R}}

\usepackage[normalem]{ulem}

\begin{document}

\title{Quasi-equilibrium models for triaxially deformed rotating compact stars}

\author{Xing Huang}
\affiliation{Department of Physics, University of Wisconsin-Milwaukee, P.O. Box 413,  
Milwaukee, WI 53201}

\author{Charalampos Markakis}
\affiliation{Department of Physics, University of Wisconsin-Milwaukee, P.O. Box 413,  
Milwaukee, WI 53201}

\author{Noriyuki Sugiyama}
\affiliation{Department of Physics, University of Wisconsin-Milwaukee, P.O. Box 413,  
Milwaukee, WI 53201}

\author{K\=oji Ury\=u}
\affiliation{Department of Physics, 
University of the Ryukyus, Senbaru, Nishihara, Okinawa 903-0213, Japan}

\date{\today}

\begin{abstract}

Quasi-equilibrium models of rapidly rotating triaxially 
deformed stars are computed in general relativistic gravity, 
assuming a conformally flat spatial geometry (Isenberg-Wilson-Mathews
formulation) and a polytropic equation of state.
Highly deformed solutions are calculated on the initial slice 
covered by spherical coordinate grids, centered at the source, 
in all angular directions up to a large truncation radius. 
Constant rest mass sequences are calculated from nearly 
axisymmetric to maximally deformed triaxial configurations. 
Selected parameters are to model (proto-) neutron stars; the compactness is $\compa = 0.001, 0.1, 0.14, 0.2$ for polytropic 
index $n=0.3$ and $\compa=0.001, 0.1, 0.12, 0.14$ for $n=0.5$. 
We confirmed that the triaxial solutions exist for these 
parameters as in the case of Newtonian polytropes. However,
it is also found that the triaxial sequences become shorter for 
higher compactness, and those may disappear at a certain large 
compactness for the $n=0.5$ case.  
In the scenario of the contraction of proto-neutron 
stars being subject to strong viscosity and rapid cooling, 
it is plausible that, once the viscosity driven 
secular instability sets in during the contraction,
the proto-neutron stars are always maximally deformed 
triaxial configurations, as long as the compactness and 
the equation of state parameters allow such 
triaxial sequences.  
Detection of gravitational waves from such sources may be 
used as another probe for the nuclear equation of state.

\end{abstract}

\maketitle

\section{Introduction}

Rapidly rotating compact objects are expected to be formed 
as new born neutron stars after stellar core collapses, or 
as differentially rotating hypermassive neutron stars after 
binary neutron star mergers.  Accretion onto neutron stars 
in X-ray binaries can also lead to rapid rotation. All of 
these have been extensively studied as 
strong sources of gravitational waves for the 
ground based laser interferometers LIGO/GEO600/VIRGO/TAMA
(See e.g. \cite{reviews} and references therein).

Classical models of rotating stars are 
a class of ellipsoidal figures of equilibrium; 
self-gravitating rotating incompressible fluids 
in Newtonian gravity.  
Such solutions include sequences of axisymmetric Maclaurin 
ellipsoids, or non-axisymmetic Jacobi, Dedekind, or Riemann 
S-type ellipsoids \cite{Ch69}. These models are 
used to study the secular evolutions of rapidly rotating 
stars due to the viscosity and the radiation back-reaction 
of gravitational waves \cite{ellipsoidal}.  
Lai and Shapiro \cite{LS95} have developed an ellipsoidal 
approximation to the rotating polytropes, and applied 
the model to clarify the secular evolution of rapidly 
rotating neutron stars in detail, and more recently focused 
on the viscosity driven secular instability \cite{STU04}.

As discussed in \cite{reviews,ellipsoidal,LS95}, 
and shown by a number of numerical simulations of 
rapidly rotating compact stars, core collapse, 
and binary neutron star mergers \cite{SimDI,SimCC,SimBNS}, 
a transient triaxially deformed compact object may survive 
within a secular time scale. In this paper, we consider uniformly 
rotating models of such triaxially deformed compact objects, 
an extension of the Jacobi ellipsoid in general relativity. 
In Newtonian gravity, such solutions exist 
for rotating polytropes with polytropic index $n < 0.808$ 
\cite{James64}.  Here, the polytropic equation of state 
(EOS) $p=\kappa\rho^{1+1/n}$ relates the pressure 
$p$ with the baryon rest mass density $\rho$.  

In general relativity, such configurations are not in equilibrium 
due to the back-reaction of gravitational radiation. However 
as in the case of quasi-equilibrium initial data of binary neutron stars, 
triaxially deformed uniformly rotating stars 
are in quasi-equilibrium, as long as the gravitational luminosity is small 
enough that the energy radiated away within a rotational period is 
small compared to the binding energy of the star, which is always the case, 
and if the viscosity is strong enough for the flow field to become uniformly 
rotating during the evolution. Therefore, as an important application, 
a sequence of uniformly rotating quasi-equilibrium 
solutions may model a secular evolution from the proto-neutron 
star to the neutron star in the strong viscosity limit, 
and each solution may serve as the initial data for the 
general relativistic hydrodynamic simulations of such objects.  

Models of rapidly rotating neutron stars have been extensively 
studied as stationary, axisymmetric, perfect-fluid spacetimes \cite{Nick03}, 
while less attention has been paid to uniformly rotating triaxial solutions, 
not only because they are not exact equilibria due to gravitational 
radiation reaction
, but also because, in early calculations \cite{FIP86}, 
 models of the EOS above nuclear density did not allow  
large enough values of $T/|W|\sim 0.14$ where a triaxial 
sequence is expected to bifurcate from an axisymmetric sequence.  
Here, $T/|W|$ is the ratio of kinetic energy $T$ to gravitational 
potential energy $W$.  However, as seen, for example, in \cite{Haensel}, 
a value $n\sim 0.5$ (an effective adiabatic index $\Gamma\sim 3$) 
may be possible for recent models of EOS for high density 
nuclear matter, above $\rho > \rho_{\rm nuc} \sim 
2\times 10^{14} {\rm g/cm}^3$, 
and, as we will see below, triaxial quasi-equilibrium 
solutions do exist even in strong gravity for relatively 
small polytropic indexes, such as $n=0.5$ or $n=0.3$, as in 
the Newtonian case.   

About a decade ago, Nozawa succeeded in computing uniformly rotating  
triaxial polytropes in general relativistic gravity 
in his thesis \cite{Nozawa}, although his calculations were 
limited by the computational resources to low resolutions.  
A few studies approximating the fluid as an ellipsoidal 
configuration in general relativistic gravity have been made
\cite{GRellpisoid}, and 
perturbative analyses have located the bifurcation point 
suggesting the existence of solutions having triaxial bar-mode 
deformations \cite{BFGall,Dorota,Saijo06,SL96,YE97,SF98}. Our computations 
of triaxially deformed stars 
can be used to locate the instability points on the axisymmetric 
sequence as well as to estimate 
the gravitational wave amplitude and luminosity from such objects. 

In this paper, we present our first results on triaxial
configurations of rapidly rotating general 
relativistic stars as models of neutron stars in extreme rotation.  
We assume a conformally flat spatial slice, and solve the constraints 
and spatial trace of the Einstein equation (Isenberg-Wilson-Mathews 
(IWM) formulation) \cite{ISEN78,WM89}
\footnote{The validity of the IWM formulation 
for axisymmetric configurations is discussed in 
\cite{CST96,WM_TEST}}.  
This is different from 
the formulation used by Nozawa \cite{Nozawa} in which the line 
element is chosen to be the same form as that of stationary 
axisymmetric spacetime, but an azimuthal dependence is allowed 
(see \cite{usui99} for the same formulation). The formulations and 
the code are described in the next section. For testing the code, 
selected axisymmetric solutions 
are compared with the results in the literature, and the 
bifurcation points of axisymmetric and triaxial sequences 
in weak gravity are examined.  Then, we present the results of 
deformation sequences of constant rest mass systematically in 
the range of two parameters, the compactness $\compa$ and 
the polytropic index $n$, appropriate for realistic neutron 
stars.  Applications of such triaxial solutions in the 
contraction of a newly born proto-neutron star are briefly 
discussed in the final section. Throughout the paper, we use units 
such that $G=c=1$.  For our tensor notation, we adopt the use of 
Greek letters for spacetime indices, and Latin letters for spatial indices.

\section{Formulation and numerical method}

\subsection{IWM formulation}
\label{sec:IWMformulation}

The IWM formulation for computing spatially conformally flat 
initial data we use for computing non-axisymmetric 
quasi-equilibrium of rotating compact star 
is briefly described.  
The spacetime ${\cal M}={\mathbb R}\times \Sigma$ 
is foliated by the family of spacelike hypersurfaces 
$\Sigma_t = \{t\}\times \Sigma$.  The future-pointing timelike 
normal $n^\alpha$ to $\Sigma_t$ is related to the timelike 
vector $t^\alpha$, which is tangent to a curve 
$t \rightarrow (t,x),\ x\in \Sigma$, by 
$t^\alpha = \alpha n^\alpha + \beta^\alpha$, where 
$\alpha$ is the lapse, and $\beta^\alpha$ the shift 
which satisfies $\beta^\alpha n_\alpha = 0$.  
Restricting the projection tensor 
$\gamma_{\albe}=\gabd + n_\alpha n_\beta$, 
a spatial metric $\gmabd (t)$ is defined on $\Sigma_t$.
In the IWM formulation, the spatial metric is chosen 
to be conformally flat, $\gmabd = \psi^4 \fabd$, 
where $\fabd$ is a flat metric on each slice, and 
$\psi$ is a conformal factor. Then the metric 
$\gabd$ takes the form 
\beq
ds^2 = -\alpha^2 dt^2
+ \psi^4 f_{ij} (dx^{i}+\beta^{i}dt) (dx^{j}+\beta^{j}dt), 
\eeq
in a chart $\{t,x^i\}$. The extrinsic curvature of the foliation is defined by 
\beq
\Kabd 
\,:=\,-\frac1{2\alpha}\pa_t \gmabd +\frac1{2\alpha}\Lie_\beta \gmabd.  
\eeq

In the IWM formulation, the Einstein equation is decomposed 
with respect to the foliation, and the following 5 components 
Eqs.(\ref{eq:Ham})-(\ref{eq:trG})
are chosen to be solved for the five metric coefficients 
$\{\psi, \alpha, \beta^a\}$ 
on the initial slice $\Sigma_0$: 
\beqn
&&(\Gabd-8\pi\Tabd)\,n^\alpha n^\beta \ \,=\, 0,
\label{eq:Ham}\\
&&(\Gabd-8\pi\Tabd)\,\gamma_a{}^\alpha n^\beta \,=\, 0,
\label{eq:Mom}\\
&&(\Gabd-8\pi\Tabd)\,\Big(\gamma^{\albe}+\frac12 n^\alpha n^\beta\Big)
\,=\, 0, 
\label{eq:trG}
\eeqn
where the first and second equations are the constraints.  
These equations are written in the form of elliptic equations 
with the non-linear source terms, respectively, 
\beqn
&&
\zLap\psi
\,=\, 
- \frac{\psi^5}{8}\left(A_{ab}A^{ab}+\frac23 K^2\right)
-2\pi\psi^5\rhoH, 
\label{eq:HaC_elip1}
\\
&&
\zLap\tbeta_a + \frac13 \zD_a\zD_b\tbeta^b
\,=\,
-2\alpha A_a{}^b\zD_b\ln\frac{\psi^6}{\alpha} 
+ \frac43\alpha\zD_aK
\nonumber\\
&&\qquad\qquad\qquad
+ 16\pi\alpha j_a,
\label{eq:MoC_elip1}
\\
&& 
\zLap(\alpha\psi)
\,=\,
-\psi^5\left(\pa_t - \Lie_{\beta}\right) K
\nonumber\\
&&\ 
+\alpha\psi^5\left(\frac{7}{8}A_{ab}A^{ab}+\frac5{12} K^2\right)
+2\pi\alpha\psi^5(\rhoH+2S),
\label{eq:trG_elip1}
\eeqn
where $K$ is the trace of $\Kabd$, $\Aabd$ its tracefree part, 
$\tbeta_a$ the conformally weighted shift defined 
by $\tbeta_a=\fabd\tbeta^b$ and $\tbeta^a=\beta^a$,
$\zLap$ is the flat Laplacian and $\zD_a$ is the covariant derivative with respect to 
the flat three-metric $\fabd$. The source terms 
of matter are defined by
$\rhoH:=\Tabd n^\alpha n^\beta$, 
$j_a:=-\Tabd \gamma_a{}^\alpha n^\beta$, and 
$S:=\Tabd \gamma^{\alpha \beta}$. 
Expressions of the sources in terms of the metric potentials 
and fluid variables are given in Appendix \ref{appsec:physq}.

We choose a maximally embedded slice $K=0=\pa_t K$.  
Because the spatial metric is conformally flat, 
$\Aabd$ does not involve time derivatives of the spatial metric, 
\beqn
\Aabd 
&=& \frac{\psi^4}{2\alpha}
\left(\Lie_\beta \fabd
- \frac13\fabd f^{cd}\Lie_\beta f_{cd}\right)
\\
&=& \frac{\psi^4}{2\alpha}
\left(\zD_a \tbeta_b + \zD_b \tbeta_a
- \frac23 \fabd \zD_c \beta^c\right)  
\eeqn
where $\Lie_\beta $ denotes the Lie derivative with respect to $\beta^a$. 
The field equations Eqs.(\ref{eq:HaC_elip1})-(\ref{eq:trG_elip1})
are thus rewritten \beqn
&& \!\!\!\!\!\!\!
\zLap\psi
\,=\, 
- \frac{\psi^5}{8}A_{ab}A^{ab}
-2\pi\psi^5\rhoH, 
\label{eq:HaC_elip2}
\\
&& \!\!\!\!\!\!\!
\zLap\tbeta_a + \frac13 \zD_a\zD_b\tbeta^b
\,=\,
-2\alpha A_a{}^b\zD_b\ln\frac{\psi^6}{\alpha} 
+ 16\pi\alpha j_a,
\label{eq:MoC_elip2}
\\
&& \!\!\!\!\!\!\!
\zLap(\alpha\psi)
\,=\,
\frac{7}{8}\alpha\psi^5A_{ab}A^{ab}
+2\pi\alpha\psi^5(\rhoH+2S).  
\label{eq:trG_elip2}
\eeqn

Eq.(\ref{eq:MoC_elip2}) is decomposed further 
to improve the accuracy 
in numerical computation. Following 
the decomposition proposed by Shibata 
\cite{shibatadecom}, we write Eq.(\ref{eq:MoC_elip2}) as 
\beq
\zLap\tbeta_a + \frac13 \zD_a\zD_b\tbeta^b = {\cal S}_a, 
\label{eq:MoC_elip3}
\eeq
and introduce 
\beq
\tbeta_a = B_a + \frac18\zD_a(B-x^bB_b),
\label{eq:decomp}
\eeq 
where 
$x^a$ are coordinates that satisfy $\zD_a x^b=\dl_a{}^b $.
Substituting the decomposition (\ref{eq:decomp}) 
into Eq.(\ref{eq:MoC_elip3}) yields 
\beq
\zLap\tbeta_a + \frac13 \zD_a\zD_b\tbeta^b
\,=\,{\zLap B_a} + \frac16\zD_a({\zLap B - x^b\zLap B_b})
\,=\, {{\cal S}_a}.
\eeq
The elliptic equations 
$\zLap B_a={\cal S}_a$ and 
$\zLap B - x^b\zLap B_b=0$ are separated, 
and the former is substituted to the latter: 
\beqn
&&
\zLap B_a \,=\, {\cal S}_a \,:=\, 
-2\alpha A_a{}^b \zD_b\ln\frac{\psi^6}{\alpha}
+ 16\pi\alpha j_a, 
\label{eq:MoC_decvec}
\\
&&
\zLap B \,=\, x^a{\cal S}_a.  
\label{eq:MoC_decsca}
\eeqn
The potentials $\{B_a, B\}$ are solved for simultaneously, 
and the shift $\tbeta_a$ is reconstructed from Eq.(\ref{eq:decomp}).  

\subsection{Formulation for a relativistic fluid in equilibrium}
\label{sec:fluidformulation}

A perfect fluid is described by the stress-energy tensor
\beq
\Tabd = (\epsilon+p)u_\alpha u_\beta+ p\gabd, \qquad 
\eeq
where $u^\alpha$ is the  4-velocity of the fluid, 
$p$ its pressure, and $\epsilon$ the energy density.
As a consequence of the Bianchi identity, the stress-energy tensor is 
covariantly conserved:
\beq 
\na_\beta \Tba=0.\label{eq:bianchiid}
\eeq 

When the fluid is close to equilibrium, 
one can obtain a simpler set of equations.  
Introducing the specific enthalpy defined by 
$h:=(\epsilon+p)/\rho$, where $\rho$ is the baryon rest 
mass density, 
the left hand side of Eq. (\ref{eq:bianchiid}) can be written
\beqn
{ \na_\beta \Tba}
&=& 
\rho\biggl[\,{ u^\beta \na_\beta(hu_\alpha) + \na_\alpha h}
\,\biggr]
\nonumber \\
&&
\,+\, hu_\alpha { \na_\beta(\rho u^\beta)}
\,-\, \rho T{ \na_\alpha s},  
\eeqn
where $s$ is the specific entropy.  
In the derivation, the local first law of thermodynamics 
$dh = Tds + dp/\rho$ was used.  
In local thermodynamic equilibrium, one can also assume 
 the conservation of baryon mass, 
\beq
\na_\alpha (\rho u^\alpha)
\,=\,
\frac{1}{\sqrt{-g}}\Lie_u (\rho\sqrt{-g})
\,=\,0.  
\eeq
Consequently, the conservation of specific entropy 
along the fluid world line, 
\beq
u^\alpha\na_\alpha s = \Lie_u s=0, 
\eeq
and, the relativistic Euler equations, 
\beq
u^\beta \na_\beta(hu_\alpha) + \na_\alpha h\,=\,
\Lie_u (hu_\alpha) + \na_\alpha h\,=\,0, 
\eeq
are obtained. Assuming the flow field to be isentropic 
everywhere inside the neutron star matter, $s=\rm const$, 
we have a one-parameter equations 
of state (EOS) $p = p(\rho)$.  

We assume a stationary state in the rotating frame 
for the fluid source. 
Imposing symmetry along the helical vector 
$k^\alpha=t^\alpha + \Omega \phi^\alpha$
where $\Omega$ is a constant angular velocity 
of a rotating frame, we have 
\beq
\Lie_{k}(\rho u^t \sqrt{-g})=0, \ \ \mbox{and}\ \ 
\gamma_a\!^{\alpha}\Lie_{k} (h u_\alpha)=0, 
\eeq with $u^t$
interpreted as the scalar $u^\alpha \na_\alpha t$. 
For a corotational flow, $ u^\alpha = u^t k^\alpha$, 
the rest mass conservation becomes trivial, 
and the relativistic Euler equation has the first integral 
\beq
\frac{h}{u^t}\,=\,{\cal E}\,=\,\mbox{constant}, 
\label{eq:firstint}
\eeq
where ${\cal E}$ is the injection energy.  
From the normalization of the four velocity $u_\alpha u^\alpha=-1$, one obtains
\beq
u^t = \frac1{\sqrt{\alpha^2 - \omega_a \omega^a}}
 = \frac1{\sqrt{\alpha^2 - \psi^4 \fabd\,\omega^a \omega^b}}, 
\label{eq:ut}
\eeq
where $\omega^a = \beta^a + \Omega \phi^a$.

As a first step in the calculation of a highly deformed 
triaxial compact star, we assume a simple polytropic EOS, 
\beq
p = \kappa \rho^{1+1/n}, 
\label{eq:polyEOS}
\eeq
where $\kappa$ is a constant, and $n$ is the polytropic 
index.  Then $h$ is related to $p/\rho$ by
\beq
h = 1 + (n+1)\frac{p}{\rho}.
\label{eq:enthalpy}
\eeq
We also refer to the polytropic exponent $\Gamma$ defined by 
$\Gamma:=1+1/n$.

\subsection{Numerical computation}

The Poisson solver and the iteration scheme used to solve the system of 
elliptic equations with non-linear source terms, are similar to 
the ones used in a previously developed initial data code 
for binary black holes and neutron stars \cite{uryu,tsokaros}. 
However, the code itself has been completely rewritten, so that 
further extensions 
can be incorporated easily. One of the revisions of the code
is that no symmetry is a priori  assumed  on the spatial slice $\Sigma_0$; 
that is, the spherical coordinate grids centered 
at the source cover all angular directions, up to a certain 
large truncation radius. Hence, for example, asymmetric magnetic fields 
may be later included 
without major modifications to the code.  
 Computation of binary solutions using the same 
coordinate setup is also possible. The other major change is a simpler, 
more robust choice of finite differencing. In this section, we briefly 
describe the necessary steps for constructing the code, which are 
1.~Spherical coordinates and the length scale, 
2.~Summary of variables and equations for coding, 
3.~Poisson solver, 4.~Grid spacing, 5.~Finite differencing and iteration, 
6.~Computation of a sequence of solutions.

\subsubsection{Spherical coordinates and the length scale}
\label{sec:coordandscale} 

The slice $\Sigma_0$ is covered by a spherical 
coordinate patch 
$(r,\theta,\phi)\in [r_a,r_b]\times [0,\pi]\times [0,2\pi]$.  
For a single star calculation, the radial coordinate 
extends from the center of the star $r = r_a = 0$ to 
the asymptotic radius $r = r_b = 10^4R_0$, 
where $R_0$ is the radius of the neutron star along 
the semi-major axis, defined by the $\theta=\pi/2$ and $\phi = 0, \pi$ lines. 
We also refer to Cartesian coordinates $(x,y,z)$ whose positive 
$x$, $y$ and $z$ directions are along $(\theta,\phi) = (\pi/2,0)$, 
$(\pi/2,\pi/2)$ and $\theta=0$, respectively.

The quantity $R_0$ is introduced as an additional parameter 
in the formulation used in our code,  
normalizing the radial coordinate as 
\beq
\hat{r} = r/R_0.  
\eeq
For a polytropic EOS, one can rescale the length 
using the polytropic constant $\kappa$    as 
$\kappa^{-n/2} R_0$, or simply setting 
$\kappa=1$  (see e.g. \cite{uryu}).  
As a result, we have three parameters 
$\{\Omega, {\cal E}, R_0\}$ in our formulation.

Furthermore, we introduce surface fitted coordinates 
on which the fluid variables are defined.  Assuming that 
the surface of a neutron star can be described by a 
function of the angular coordinate 
$R(\theta,\phi) = R_0 \hat{R}(\theta,\phi)$, 
the surface fitted coordinates $(\hat{r}_{\rm f},\theta_{\rm f},
\phi_{\rm f})$ are defined by 
\beq
\hat{r}_{\rm f}:=\hat{r}/\hat{R}(\theta,\phi),\ \ 
\theta_{\rm f} = \theta,\ \ \phi_{\rm f}  = \phi, 
\eeq
where $\hat{r}_{\rm f}$ is defined in a region $\hat{r}_{\rm f} \in [0,1]$.

\subsubsection{Summary of variables and equations for coding}

As mentioned in Sec. \ref{sec:IWMformulation},  
the field equations (\ref{eq:HaC_elip2}), (\ref{eq:trG_elip2}),
(\ref{eq:MoC_decvec}) and (\ref{eq:MoC_decsca})
are solved for the metric potentials $\{\psi,\alpha\psi,B^a,B\}$
and, as in Sec. \ref{sec:fluidformulation} and \ref{sec:coordandscale},
a comoving fluid in equilibrium is characterized by 
one fluid variable, which is chosen to be the relativistic enthalpy 
$\{h\}$, and  three parameters $\{\Omega,{\cal E},R_0\}$.  

The field equations are normalized to have the following form; 
representing each of the metric potentials $\{\psi,\alpha\psi,B^a,B\}$ 
by $\Phi$, 
\beq
\zLap \Phi \,=\, 
S_{\rm g} \,+\, R_0^2\, S_{\rm m}, 
\label{eq:elip}
\eeq
where the flat Laplacian $\zLap$  corresponds now to the normalized 
coordinate $\hat{r}$.  
The source term $S_{\rm g}$ includes the metric potentials 
and their derivatives, while $S_{\rm m}$ also includes the matter 
variables and the parameters $\{\Omega, \cal E\}$, while 
the dependence on the length scale $R_0$ is explicitly 
separated in Eq.(\ref{eq:elip}).

The fluid variable $\{h\}$ is determined by 
Eq.(\ref{eq:firstint}) coupled to the EOS (\ref{eq:polyEOS}), and 
the relations (\ref{eq:ut}) and (\ref{eq:enthalpy}). The 
three parameters $\{\Omega,{\cal E},R_0\}$ are  determined by the  following 
three quantities: the surface radii along two of the three 
semi-major axes, and the value of the central density.
These quantities are used to impose three conditions on Eq.(\ref{eq:firstint}), 
which are solved with respect to 
the three parameters $\{\Omega, {\cal E}, R_0\}$
in each iteration cycle.  

\subsubsection{Poisson solver}
The elliptic equations (\ref{eq:HaC_elip2}), (\ref{eq:trG_elip2})
(\ref{eq:MoC_decvec}), and (\ref{eq:MoC_decsca})
are integrated on the spherical grid using Green's formula.  
Representing each of the potentials $\{\psi,\alpha\psi,B^a,B\}$ 
by $\Phi$, the latter is given by 
\beqn
\Phi(x)&=& -\frac1{4\pi}\int_{V} G(x,x')S(x') d^{3}x' 
\nonumber \\
&&
+ \frac{1}{4\pi} \int_{\pa V} \left[G(x,x')\na'^a \Phi(x')\right.
\nonumber \\
&&\qquad\qquad
\left. - \Phi(x')\na'^a G(x,x') \right]dS'_a.  
\label{eq:GreenIde}
\eeqn
where $x$ and $x'$ are positions, $x,x'\in V \subseteq \Sigma_0$.
We choose the Green function $G(x,x')$ without boundary, 
\beq
\zLap G(x,x') = -4\pi\delta(x-x'), 
\eeq
and perform a multipole expansion in associated Legendre functions,  
\beqn
&&
G(x,x')=\frac{1}{\left|{x}-{x'}\right|}\,=\, 
\sum_{\ell=0}^\infty g_\ell(r,r') \sum_{m=0}^\ell \epsilon_m \,
\frac{(\ell-m)!}{(\ell+m)!}
\nonumber\\
&&\qquad\quad
\times
P_\ell^{~m}(\cos\theta)\,P_\ell^{~m}(\cos\theta')
\cos m(\varphi-\varphi'), 
\eeqn
where the radial Green function $g_\ell(r,r')$ is defined by 
\beq
g_\ell(r,r')=\frac{r_<^\ell}{r_>^{\ell+1}}, 
\quad r_> := \sup\{r,r'\}, \ r_< := \inf\{r,r'\}, 
\eeq
and the coefficients $\epsilon_m$ are equal to $\epsilon_0 = 1$ for $m=0$, 
and $\epsilon_m = 2$ for $m\ge 1$.

\subsubsection{Grid spacing}

The field equations in the integral form (\ref{eq:GreenIde}) 
are discretized on the spherical grids, and iterated until convergence is achieved.  
Our code allows us to use any non-equidistant 
grid spacing in all the spatial coordinates, 
$(r_i,\theta_j,\phi_k)$, 
$i = 0,\cdots, N_r$, $j = 0,\cdots, N_\theta$, and 
$k = 0,\cdots, N_\phi$. 
The radial grid points are equidistant in the region 
$[r_a,r_c]$ and non-equidistant in $[r_c,r_b],$ as follows:
writing $\Dl r_i := r_i - r_{i-1}$, we have
\beqn
&&
\Dl r_i=\Dl r=\frac{r_c-r_a}{n_r} 
\quad \mbox{for}\quad   i = 1, \cdots, n_{r}, 
\\
&&
\Dl r_i = k \Delta r_{i-1}  
\qquad \mbox{for}\quad  i = n_r+1, \cdots,  N_r, 
\eeqn
where the constant $k$ is determined from the relation 
\beq
r_{b}-r_{c}=\frac{k^{N_{r}-n_{r}+1}-k}{k-1}\Dl r  \:.
\eeq
We choose equidistant grid spacing for 
$\theta_j$ and $\phi_k$, that is, 
$\Dl \theta_i = \Dl \theta = \pi/N_\theta$, 
and $\Dl \phi_i = \Dl \phi = 2\pi/N_\phi$.  
Our notations for the grid points are summarized in 
Table \ref{tab:grid}.  
\begin{table}
\begin{tabular}{lll}
\hline
$r_{a}$ &:& Radial coordinate where the grid $r_i$ starts.                  \\
$r_{b}$ &:& Radial coordinate where the grid $r_i$ ends.                \\
$r_{c}$ &:& Radial coordinate between $r_{a}$ and $r_{b}$ where the      \\
&\phantom{:}& grid changes from equidistant to non-equidistant.          \\
$N_{r}$ &:& Total number of intervals $\Dl r_i$ between $r_{a}$ and $r_{b}$. \\
$n_{r}$ &:& Number of intervals $\Dl r_i$ between $r_{a}$ and $r_{c}$.       \\
$N_{\hat{r}}$ &:& Total number of intervals $\Dl \hat{r}_i$ for 
$\hat{r}\in[\hat{r}_a,\hat{R}(\theta,\phi)]$. \\
$N_{\theta}$ &:& Total number of intervals $\Dl \theta_i$ for $\theta\in[0,\pi]$. \\
$N_{\phi}$ &:& Total number of intervals $\Dl \phi_i$ for $\phi\in[0,2\pi]$. \\
\hline
\end{tabular}  
\caption{Summary of grid parameters.}
\label{tab:grid}
\end{table}

\subsubsection{Finite differencing and iteration}
\label{sec:FDandIter}

For the numerical integration of Eq.(\ref{eq:GreenIde})
we select the mid-point rule. Accordingly, source 
terms are evaluated at the middle of successive 
grid points. The linear interpolation formula and 
the second order Lagrange formula are applied for computing the source term 
fields and their derivatives respectively, at the mid-points of the 
$r$, $\theta$ and $\phi$ grids.  

The reason for selecting a rather low (second) order 
finite difference scheme is the following: When 
the field quantities vary rapidly, such as at a density 
discontinuity in a neutron star, higher order interpolating 
formulas as well as finite difference formulas tend to 
overshoot near the region, and may cause a non-convergent 
iteration.  To overcome this behavior, one may either 
(i) separate the computing regions at the discontinuity, 
or (ii) use lower order polynomial approximations.  
With the first approach, pseudo-spectral methods have been 
 successfully implemented by \cite{Meudon} and achieved 
an evanescent error.  We select the second idea to keep 
the code as simple and flexible as possible, and improve the  accuracy by 
simply increasing the number of grid points.  

In each iteration cycle, the Poisson solver (\ref{eq:GreenIde})
is called for each variable.  Writing the L.H.S. of 
Eq.(\ref{eq:GreenIde}) as $\hat{\Phi}$, 
each field variable is updated from the $N$th iteration 
cycle to the $(N+1)$th in the manner 
\beq
\Phi^{(N+1)} \,=\, \lambda\hat{\Phi} \,+\, (1-\lambda)\Phi^{(N)}, 
\eeq
where the softening parameter $\lambda$ is chosen to be 
$0.3 \sim 0.5$ for accelerated convergence. Then we check the relative 
difference of successive cycles
\beq
\frac{2\,|\,\Phi^{(N+1)}  \,-\, \Phi^{(N)}\,|}
{|\,\Phi^{(N+1)}\,|  \,+\, |\,\Phi^{(N)}\,|}, 
\eeq
as a criteria for the convergence. 
We typically stop the iteration when 
this quantity becomes less than $10^{-6}$.   

The method used in this code may be considered as 
an extension of the one developed by Ostriker and Marck (1968)
\cite{OM68} for Newtonian rotating stars, and by 
Komatsu, Eriguchi, and Hachisu (1989) for relativistic 
rotating (axisymmetric) neutron stars, known as the KEH code \cite{KEH89}.

\subsubsection{Constructing a sequence of solutions}

We compute constant rest mass sequences of isentropic equilibrium solutions 
in the IWM formulation.   Constant entropy is modeled by 
setting the parameter $\kappa$ in the EOS to a constant.  
In a quasi-equilibrium evolution of a rotating star, 
the angular velocity remains constant when the viscosity  
of matter is dominant. Then the solution sequence approximately
models an evolution, with an error that includes neglecting the increase 
of entropy due to viscosity.  

Each solution is computed by setting the central value of 
$q := p/\rho$ to $q = q_c$, 
the ratio $R(0,\phi)/R_0$ of the stellar radii along the $z$ and $x$ 
axes for the axisymmetric configuration, 
and the ratio $R(\pi/2,\pi/2)/R_0$ of the $y$ and $x$ axes for the 
triaxial configuration.
To compute a constant rest mass sequence, one iterates 
$M_0(q_c)$ changing $q_c$ until $M_0$ converges to the specified 
value. We use a discrete Newton-Raphson iteration for the rest mass.  

A sequence of rotating star solutions with a certain EOS is labeled by  
the compactness $\compa$ of a non-rotating 
spherical star having the same rest mass $M_0$. We denote the gravitational 
mass of this spherical star by $M$, and the compactness by $\compa$.  
This labeling for each sequence 
is possible as long as it is a normal sequence
that has a stable spherical star in the limit that 
the angular velocity $\Omega$ goes to zero, 
which is not the case for a supermassive sequence.  
In the following, we focus on the normal sequences 
whose compactness is close to its value 
for a neutron star, around $\compa\sim0.1-0.2$.

Formulas used for computing the rest mass $M_0$, 
ADM mass $\Madm$, Komar mass $\MK$, as well as the 
total angular momentum $J$ are presented in 
Appendix \ref{appsec:physq}.  
For polytropic EOS, one can normalize these 
quantities by a certain power of the polytropic 
constant $\kappa$, as shown in the same Appendix.  
Hence we choose $\kappa=1$ units to 
present solution sequences.  

A sequence of solutions with constant rest mass is considered as 
an evolutionary track of adiabatic changes in quasi-equilibrium. 
Under this assumption, the solutions in each sequence 
are parameterized by the angular velocity $\Omega$, 
and the first-law relation 
\beq
\dl\Madm = \Omega \dl J
\eeq
is satisfied, as proved in \cite{FUS}.

\section{Code test}

\subsection{Axisymmetric solutions}

Axisymmetric solutions calculated by our new code 
are compared with the results in the literature 
\cite{CST96,NSGE98}.  
We show the results of comparisons for models presented 
in Table I of Cook,\ Shapiro and Teukolsky \cite{CST96} (hereafter CST), 
which correspond to a solution sequence with constant 
rest mass $M_0 = 0.14840$ for the case with 
the polytropic index $n=0.5$.  
This value of $M_0$ is close to the maximum 
rest mass of a non-rotating spherical solution; 
the gravitational mass and the compactness 
of the same non-rotating solution are 
$M=0.12304$ and $\compa = 0.29605$. 
In Table \ref{tab:CSTcompared}, 
selected solutions calculated with the highest 
resolution I-5 in Table \ref{tab:reso} 
are compared with the results shown in Table I of 
\cite{CST96}.  Fractional errors in any quantities 
are less than $0.5\%$.

\begin{table}
\begin{tabular}{ccccccccccc}
\hline
Type&$r_a$&$r_b$&$r_c$&$N_r$&$n_r$&$N_{\hat{r}}$&
$N_\theta$&$N_\phi$&$L$ \\
\hline
I-1 &0&$10^4$&1.25& 60& 20& 16& 24& 48& 12\\
I-2 &0&$10^4$&1.25& 90& 30& 24& 36& 72& 12\\
I-3 &0&$10^4$&1.25&120& 40& 32& 48& 96& 12\\
I-4 &0&$10^4$&1.25&180& 60& 48& 72&144& 12\\
I-5 &0&$10^4$&1.25&240& 80& 64& 96&192& 12\\
\hline
II-1 &0&$10^4$&1.25&120& 40& 32& 24& 48& 8\\
II-2 &0&$10^4$&1.25&180& 60& 48& 36& 72& 10\\
II-3 &0&$10^4$&1.25&240& 80& 64& 48& 96& 12\\
\hline
\end{tabular}
\caption{Coordinate parameters, and the number of grid points 
with different resolutions.  
$L$ is the highest multipole included in the Legendre expansion.}
\label{tab:reso}
\end{table}
\begin{table}
\begin{tabular}{ccccccccccc}
\hline
&$e$&$\Omega$&$\Madm$&$T/|W|$&$\epsilon_{\rm{c}}$\\
\hline
Present&0.4614&0.5252&0.1247&0.04281&0.7911\\
CST&0.4592&0.5232&0.1247&0.04253&0.7911\\
\hline
Present&0.6370&0.6672&0.1264&0.08711&0.6613\\
CST&0.6360&0.6658&0.1266&0.08705&0.6613\\
\hline
Present&0.7581&0.7222&0.1281&0.1310&0.5614\\
CST&0.7585&0.7214&0.1284&0.1314&0.5614\\
\hline\\
\end{tabular}
\caption{The numerically obtained values are compared with those 
based on Table 1 of CST.  Model parameters of these solutions are
$n=0.5$, $\compa=0.298$, $M_0 = 1.484\times 10^{-1}$, and 
$M =  1.230\times 10^{-1}$.  
The quantities from CST are interpolated to have the 
same central energy density $\epsilon_{\rm{c}}$ using the four-point 
Lagrange interpolating polynomials.  
The gravitational potential energy $W$ is defined by 
Eq.~(\ref{eq:grav_pot_en}).  The eccentricity 
$e$ is defined by $e:=\sqrt{1-(\bR_z/\bR_x)^2}$ where 
the radii $\bR_x, \bR_z$ along the 
$x$ and $z$ axes are measured in proper length as in 
Eq.~(\ref{eq:proper_length}).  
}
\label{tab:CSTcompared}
\end{table}

As discussed in the section \ref{sec:FDandIter}, our choice of 
finite difference approximations is second order.  The rate 
of convergence of our code is checked using different resolutions, 
whose setups of coordinate grids are shown in Table \ref{tab:reso}.  
The grid spacing of each coordinate $(\Delta r, \Delta \theta, \Delta \phi)$
is proportionally scaled as $2/3$, $3/4$, $2/3$, $3/4$, 
from type I-1 to I-5.  
Here, we show the results of the convergence test with respect to 
the resolutions, fixing the maximum number of multipoles 
$L$ as shown in Table \ref{tab:reso}.  
\footnote{For the convergence tests with respect to 
the order of the Legendre expansion, see \cite{tsokaros}}.  

\begin{figure}
\begin{center}
\includegraphics[height=60mm]{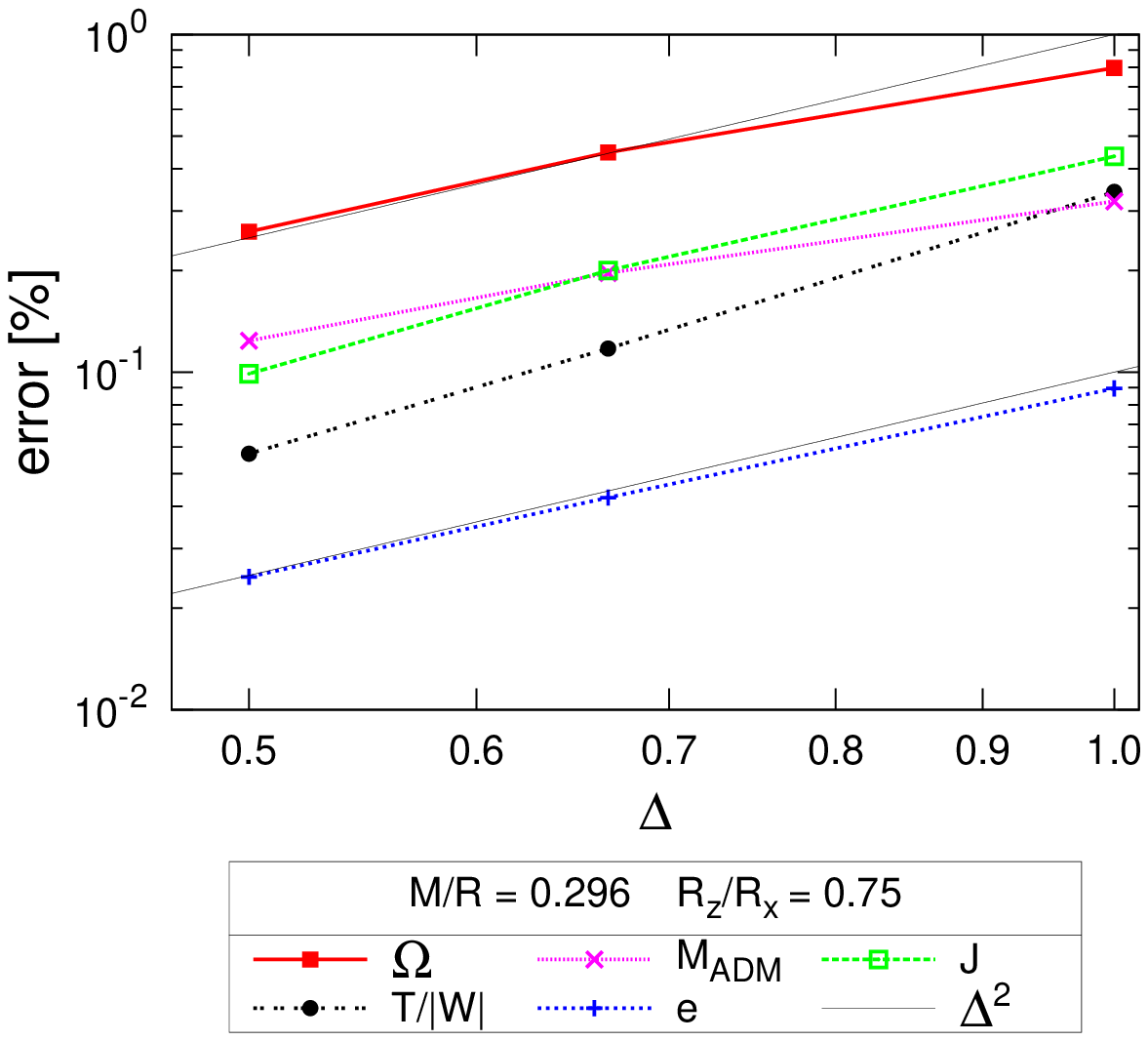}
\end{center}
\caption{The convergence of quantities, $\Omega$, $\Madm$, $J$, $T/|W|$ 
and $e:=\sqrt{1-(\bR_z/\bR_x)^2}$ (in the proper length) for the model 
with $n=0.5$, $\compa=0.296$ and axis ratio in the coordinate length 
$R_z/R_x = 0.75$.  
Normalized differences 
$|(f_{{\rm I}\mbox{-}5}-f_{{\rm I}\mbox{-}3})/f_{{\rm I}\mbox{-}5}|$, 
$|(f_{{\rm I}\mbox{-}4}-f_{{\rm I}\mbox{-}2})/f_{{\rm I}\mbox{-}5}|$, and
$|(f_{{\rm I}\mbox{-}3}-f_{{\rm I}\mbox{-}1})/f_{{\rm I}\mbox{-}5}|$
discussed in the text are plotted from left to right for each quantity 
against the resolutions $\Delta_{{\rm I}\mbox{-}3}$, 
$\Delta_{{\rm I}\mbox{-}2}$, and $\Delta_{{\rm I}\mbox{-}1}$, 
respectively.  
Black thin lines are proportional to $\Delta^2$.}
\label{fig:convtest}
\end{figure}

When a sufficient number of multipoles is kept, 
the differences between numerically computed quantities 
with different resolutions and their exact value 
are written 
\beq
f_{{\rm I}\mbox{-}i}\,-\,f_{\rm exact}
\,=\,A\Delta^n_{{\rm I}\mbox{-}i} + \Od(\Delta^{n+1}_{{\rm I}\mbox{-}i})
\eeq 
where $f_{{\rm I}\mbox{-}i}$ ($i = 1, \cdots, 5$) 
is a quantity computed using one of 
the resolution types ${\rm I}\mbox{-}i$ in Table \ref{tab:reso},
 $f_{\rm exact}$ is its exact value, 
$\Delta_{{\rm I}\mbox{-}i}$ represents the grid spacing associated with 
the type ${\rm I}\mbox{-}i$ setup, and $A$ is a constant.  
Then, keeping the leading term, differences between different 
resolutions become 
\beq
f_{{\rm I}\mbox{-}k}\,-\,f_{{\rm I}\mbox{-}i}
\,=\,
A\left[
\left(\frac{\Delta_{{\rm I}\mbox{-}k}}{\Delta_{{\rm I}\mbox{-}i}}\right)^n-1
\right]
\Delta^n_{{\rm I}\mbox{-}i}.  
\label{eq:convtest}
\eeq
To see the order $n$ in a log-log plot, we select the combinations 
of different resolutions that give the same ratio 
$\Delta_{{\rm I}\mbox{-}k}/\Delta_{{\rm I}\mbox{-}i}$, and 
in our choice, these are 
$f_{{\rm I}\mbox{-}3}-f_{{\rm I}\mbox{-}1}$,  
$f_{{\rm I}\mbox{-}4}-f_{{\rm I}\mbox{-}2}$, and 
$f_{{\rm I}\mbox{-}5}-f_{{\rm I}\mbox{-}3}$. 
In Fig.\ref{fig:convtest}, these combinations normalized 
by $f_{{\rm I}\mbox{-}5}$ of 
selected quantities are plotted against the grid spacing, 
where $\Delta$ represents the grid spacing in arbitrary units. 
It is clearly seen that the local quantities, here  
$\Omega$ and $e$, converge to $\Od(\Delta^2)$, 
and that integral (global) quantities also
approach second order convergence as the resolution 
increases.  

We also checked the convergence with the different sets of resolutions 
type II-1, 2 and 3.  These setups have fewer grid points in the angular 
coordinates $\theta$ and $\phi$.  We found that the highest resolution 
type II-3 agrees well with the results of the higher resolutions 
of type I-4 or 5 for the axisymmetric solutions.

\begin{figure}
\begin{center}
\includegraphics[height=70mm,clip]{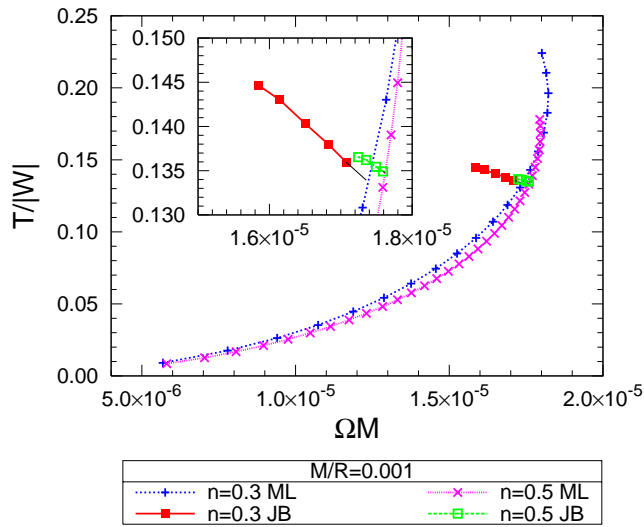}
\end{center}
\caption{Plot of $T/|W|$ versus normalized angular velocity $\Omega M$ 
for triaxial solution sequences (labeled by JB) are shown with curves 
marked with filled squares for $n=0.3$, and with squares for $n=0.5$.  
These sequences merge with axisymmetric solution sequences (labeled 
by ML) of the corresponding parameters (inset for a close up), which 
are shown by curves marked with plus (+) for $n=0.3$, and with 
crosses $(\times)$ for $n=0.5$.  The compactness $\compa$ of 
the sequences is set $\compa=0.001$ for modeling the weak 
gravity regime.}
\label{fig:ToverW_C0001}
\end{figure}

\subsection{Triaxial solutions with $\compa = 0.001$}

We calculate triaxial sequences for 
 small compactness $\compa = 0.001$, which is in 
the Newtonian regime, to check the value of $T/|W|$ 
at the bifurcation point of the triaxial sequence 
from the axisymmetric sequence.  
The triaxial and axisymmetric solution sequences for 
$n=0.3$ and $0.5$ are plotted in Fig.~\ref{fig:ToverW_C0001}. 
Extrapolating the triaxial sequences to 
corresponding axisymmetric sequences, values at the branch 
points are determined approximately as 
$(\Omega M, T/|W|)=(1.735\times 10^{-5},0.134)$ and 
$(1.763\times 10^{-5},0.135)$ for $n=0.3$ and $0.5$, 
respectively.  This value may be compared with the
Newtonian results such as the ellipsoidal approximation  
$T/|W| = 0.138$ for $n=0.5$ \cite{STU04}.  

\section{Triaxial solutions}

\subsection{Accuracy of the sequences of solutions}

Triaxially deformed solutions are calculated for selected 
values of the polytropic index, $n = 0.3$ and $0.5$. 
 Models with $\compa=0.001, 0.1, 0.14, 0.2$ are calculated 
for $n=0.3$ and with $\compa=0.1, 0.12, 0.14$ for $n=0.5$.  

We noticed that it is necessary to increase the numbers of 
grid points as much as in type I-5 in Table \ref{tab:reso} 
to have a smoothly changing sequence of triaxial solutions.  
For lower resolutions, the sequences appear 
to be less smooth especially for the plot of $T/|W|$ 
and for the part of the sequences closer to axisymmetric 
solutions. One of the reasons for this may be that, 
when one compares neighboring solutions of 
 deformed sequences for mass, binding energy, 
or angular momentum, the change in these 
quantities for triaxial sequences is much 
smaller 
than that for axially symmetric sequences of about 
the same amount of deformation.  

\begin{figure}
\begin{tabular}{cc}
\begin{minipage}{.5\hsize}
\begin{center}
\includegraphics[height=35mm,clip]{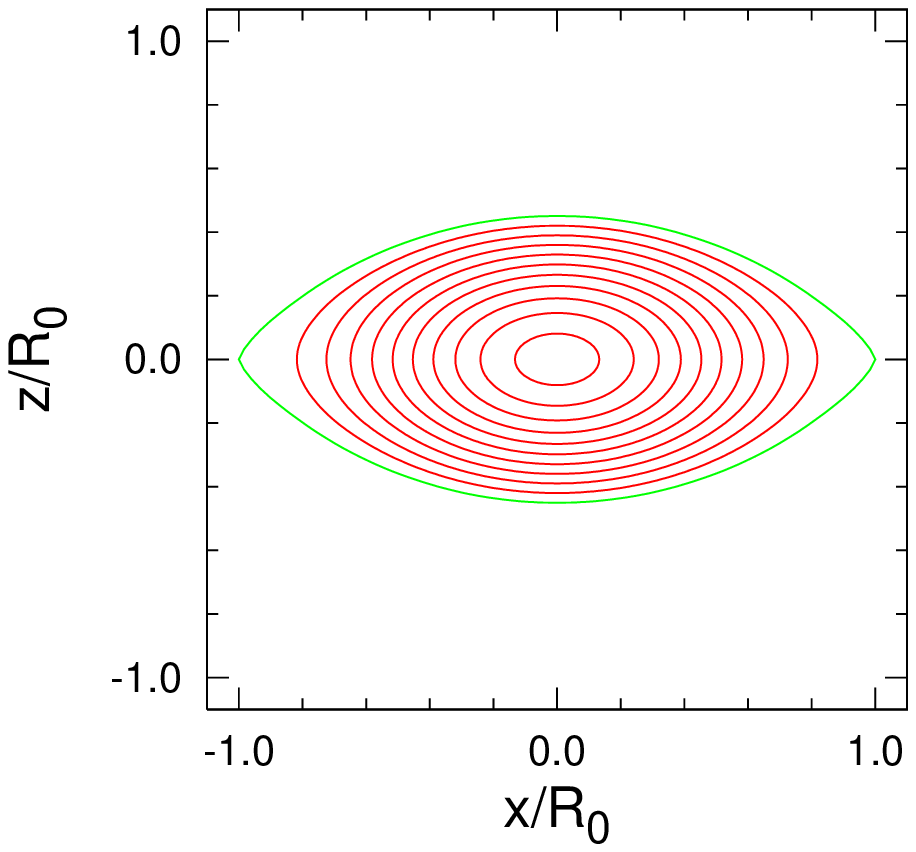}
\end{center}
\end{minipage} 
&
\begin{minipage}{.5\hsize}
\begin{center}
\includegraphics[height=35mm,clip]{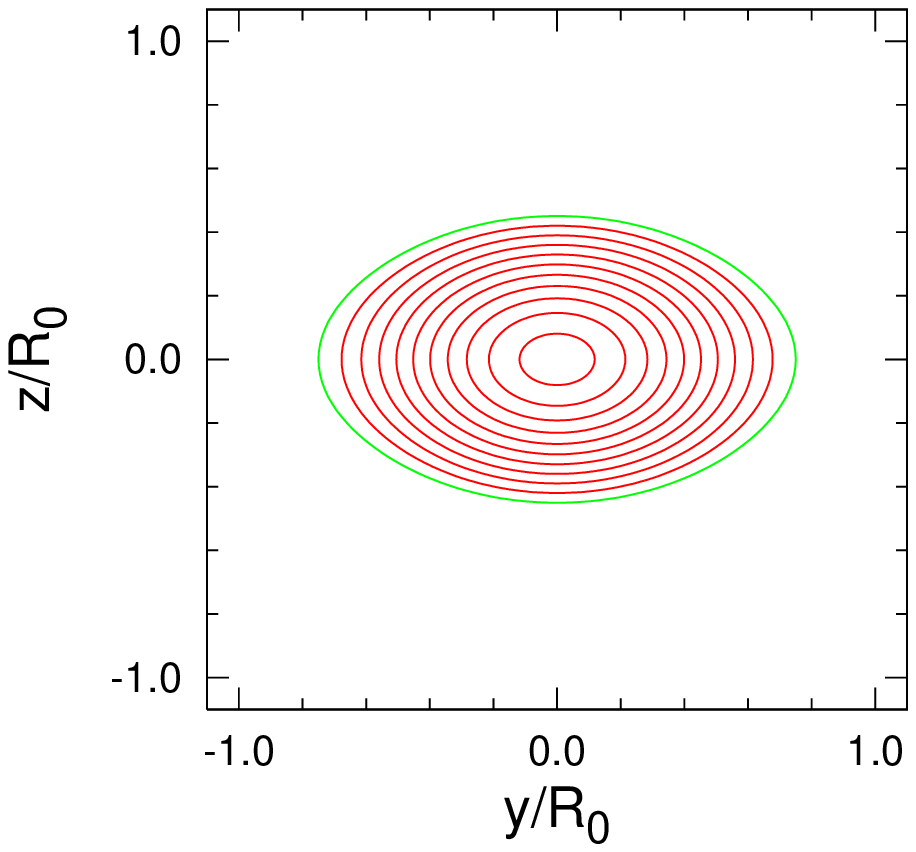}
\end{center}
\end{minipage} 
\\
\begin{minipage}{.5\hsize}
\begin{center}
\includegraphics[height=35mm,clip]{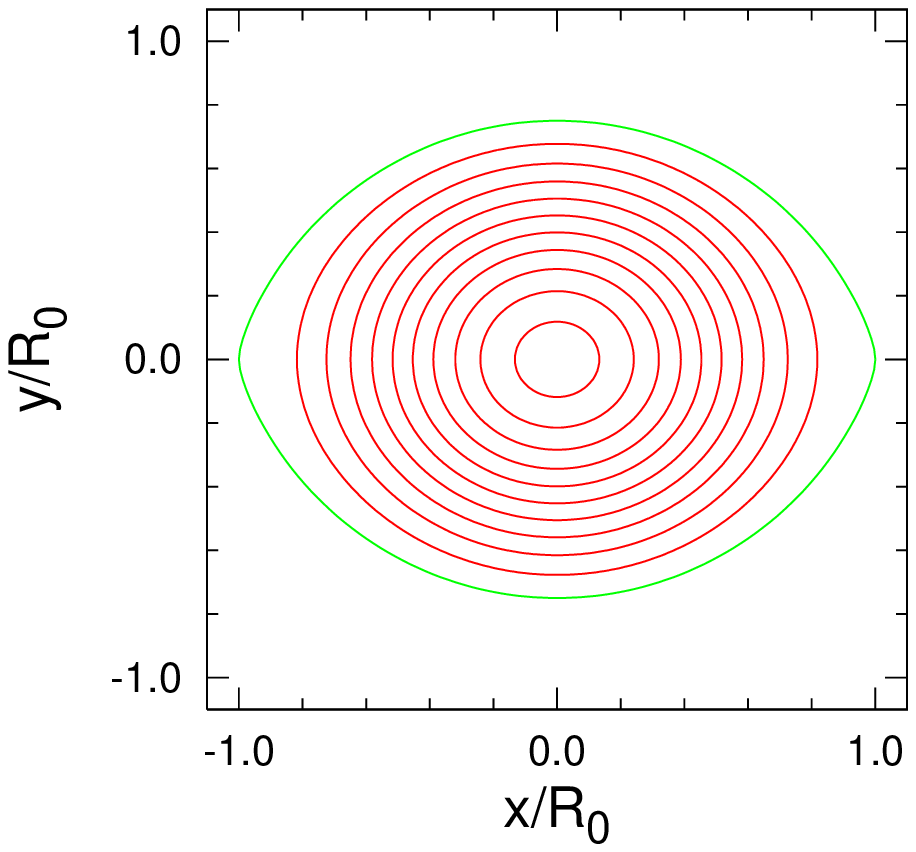}
\end{center}
\end{minipage} 
&
\begin{minipage}{.5\hsize}
\begin{center}
\includegraphics[height=20mm,clip]{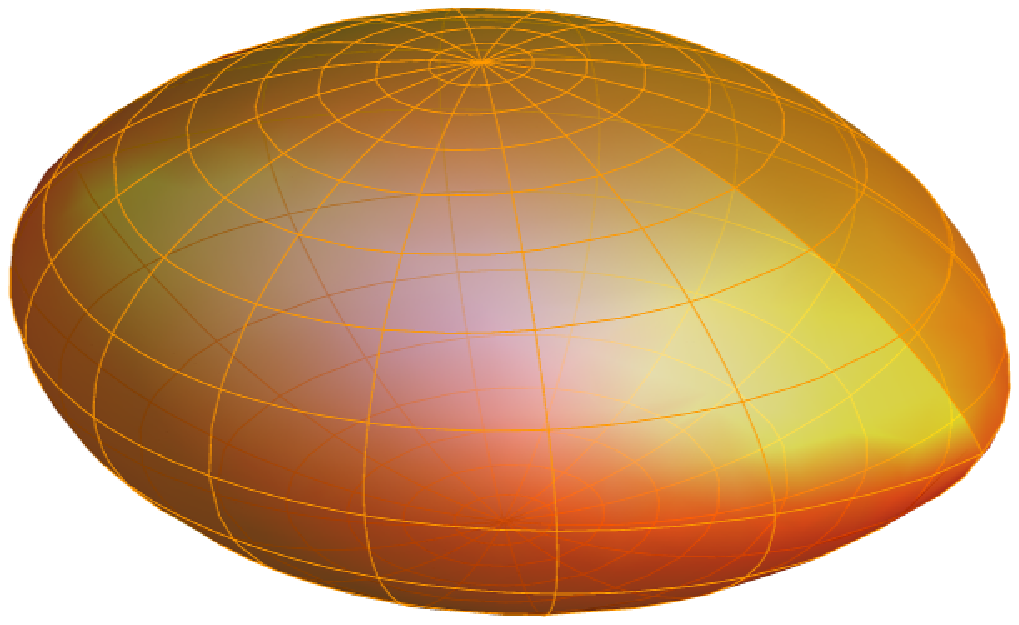}
\end{center}
\end{minipage} 
\end{tabular} 
\caption{Contours of the $p/\rho$ on $xz$-plane (top left panel), 
on $yz$-plane (top right panel), and on $xy$-plane (bottom left panel) 
are shown for the most deformed  triaxial model of $n=0.3$ 
and $\compa=0.2$.  Contours are drawn linearly from 0.0 to 0.1 every 
0.01 step.}  
\label{fig:shapes}
\end{figure}
\begin{figure}
\begin{center}
\includegraphics[height=70mm]{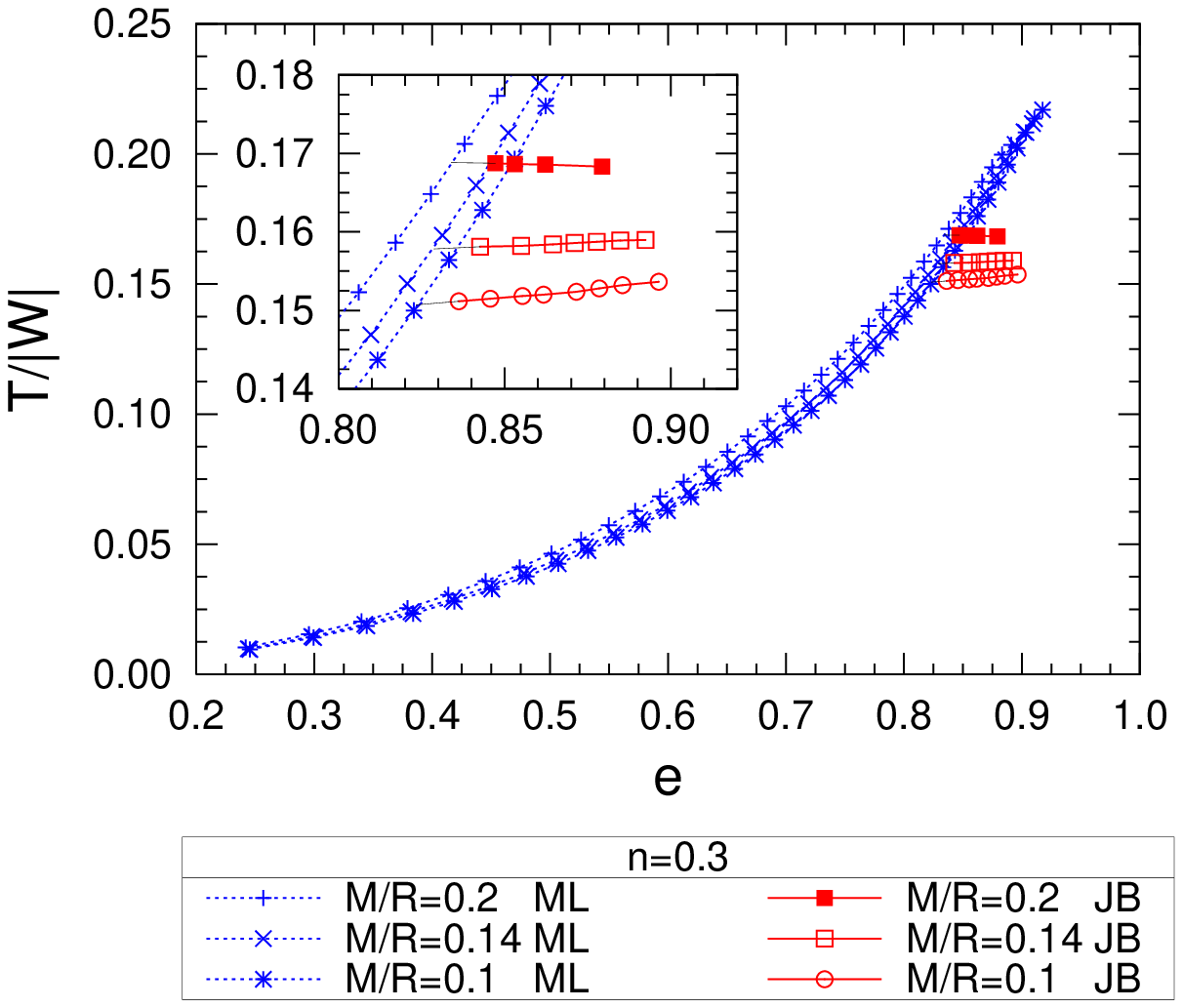}
\\[5mm]
\includegraphics[height=70mm]{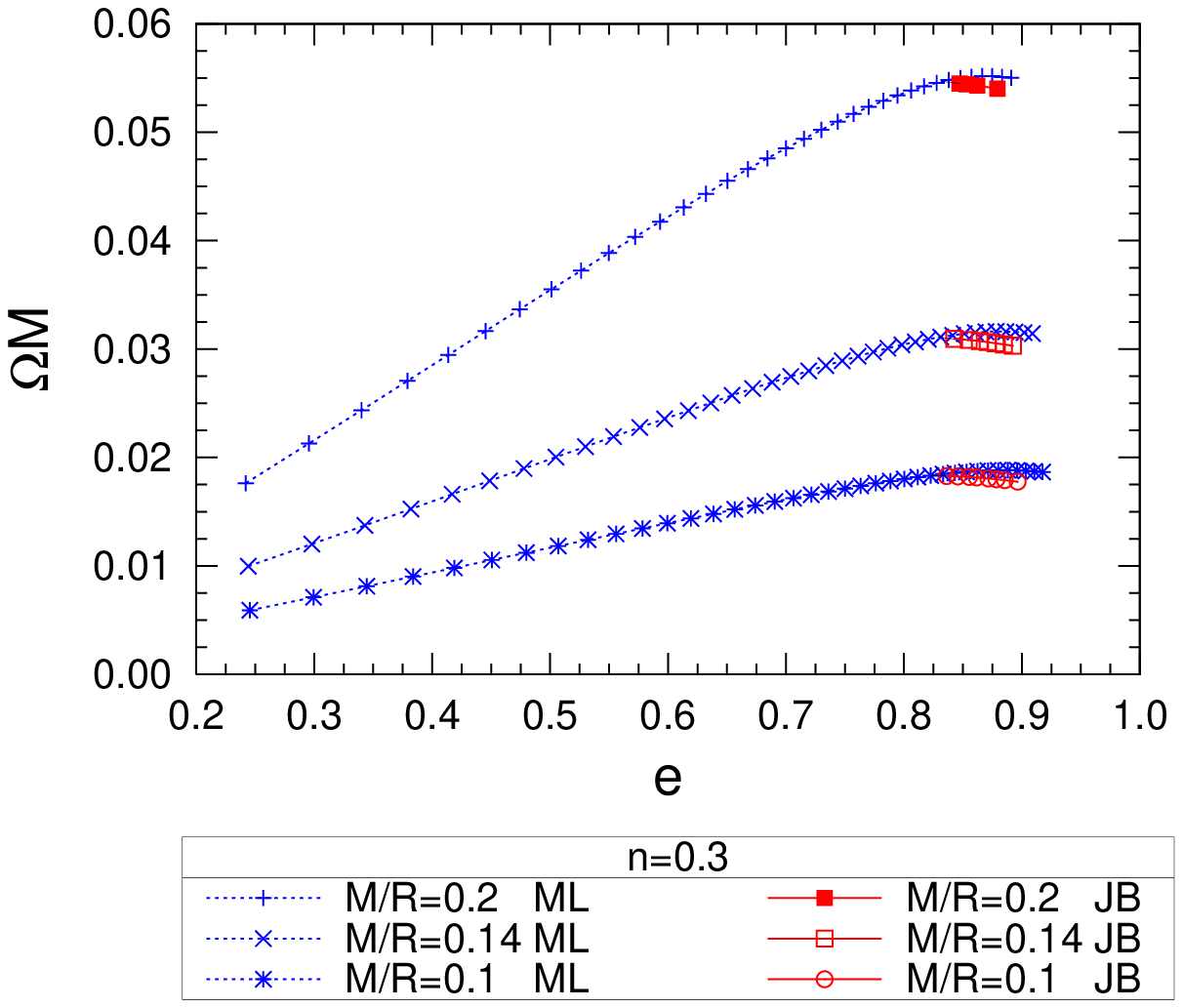}
\end{center}
\caption{Plots for $T/|W|$ (top panel) and $\Omega M$ (bottom panel) 
versus eccentricity $e:=\sqrt{1-(\bR_z/\bR_x)^2}$ (in proper length) for 
$n=0.3$ sequences.  Dashed curves labeled ML are axisymmetric 
solution sequences, and solid curves labeled JB 
triaxial solution sequences, where those correspond, 
from the top curves to the bottom in each panel, to
$\compa=0.2$, $0.14$ and $0.1$ respectively.
}
\label{fig:plot_n03}
\end{figure}
\begin{figure}
\begin{center}
\includegraphics[height=70mm]{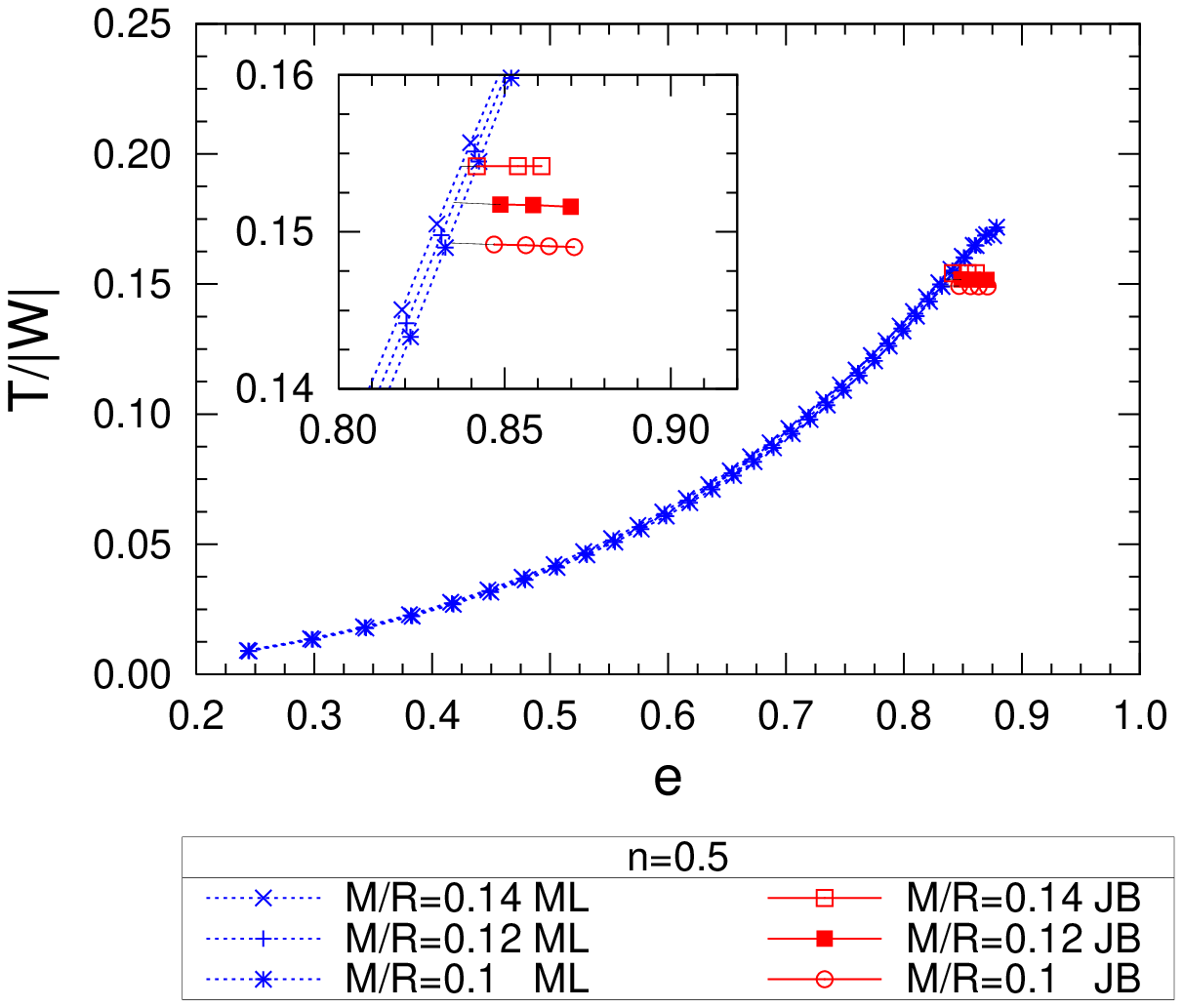}
\\[5mm]
\includegraphics[height=70mm]{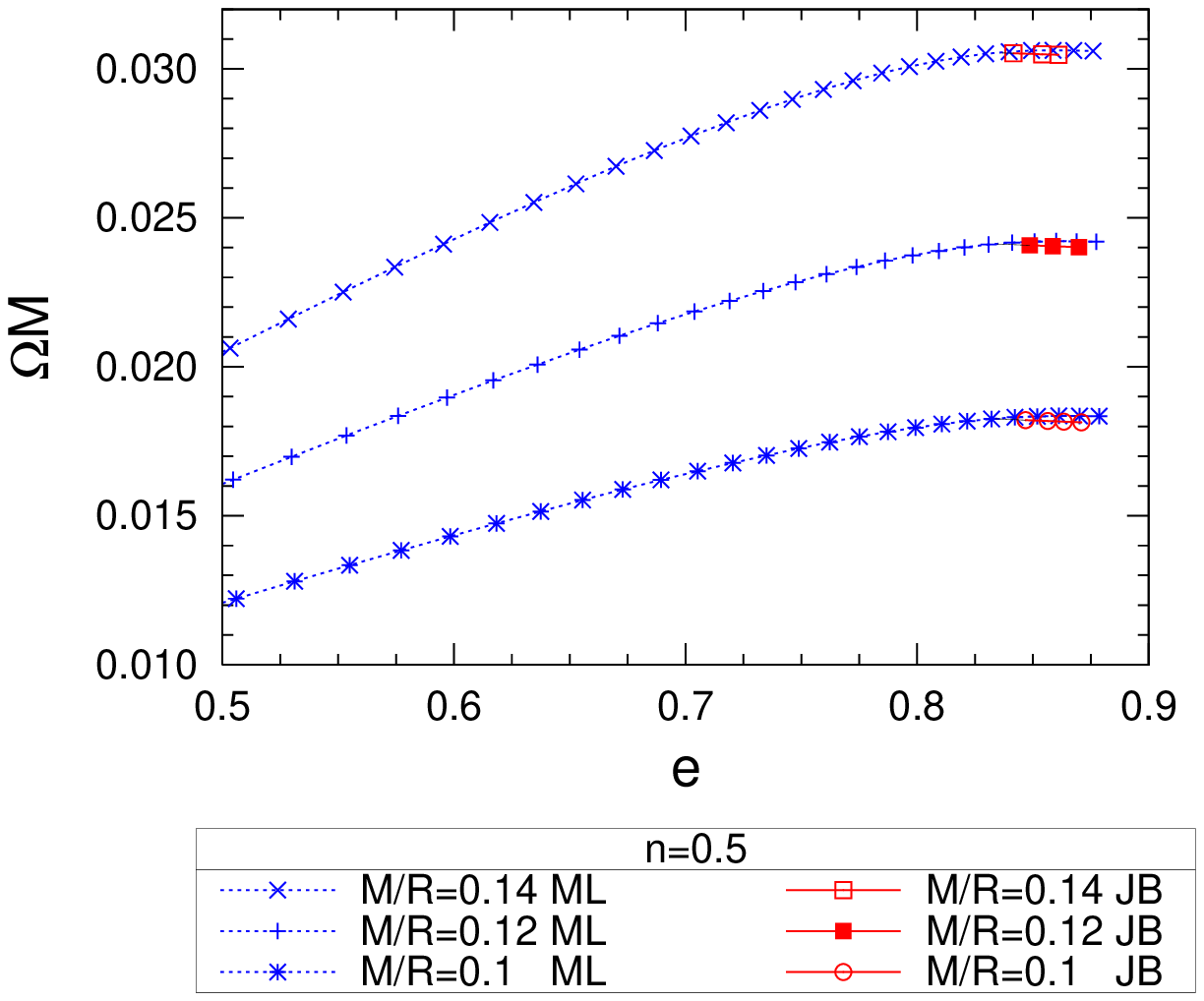}
\end{center}
\caption{Same as Fig.~\ref{fig:plot_n03} but for $n=0.5$ sequences.
Dashed curves and solid curves from the top to the bottom 
in each panel correspond to $\compa=0.14$, $0.12$ and $0.1$ 
respectively.
}
\label{fig:plot_n05}
\end{figure}

\subsection{Properties of the triaxial sequences}

As a sample of calculated solutions, the density contours in the 
$xy$, $xz$, and $yz$ planes and the surface plot of the model 
with parameters $n=0.5$ and $\compa=0.2$ and the largest deformation 
are presented in Fig.\ref{fig:shapes}.  
The solution corresponds to the last row of data 
shown in Table \ref{tab:n03n05all} in Appendix \ref{appsec:seq}.  

In Figs.~\ref{fig:plot_n03} and \ref{fig:plot_n05}, 
$T/|W|$ and $\Omega M$ are plotted for $n=0.3$ and $0.5$ 
respectively, against the eccentricity for the constant 
rest mass sequences shown in the same Table \ref{tab:n03n05all}.  
For uniformly rotating Newtonian polytropes, 
$T/|W|$ at the bifurcation point weakly depends on 
the difference of the EOS parameter, 
the polytropic index, whose value is about $T/|W|\sim 0.14$.  
For highly differential rotations, $T/|W|$ may vary largely
\cite{Saijo06,Yoshida02}.  The definitions of $T$ and $W$ 
in general relativity are given in Appendix \ref{appsec:physq}.  

Our results for the solution sequences of uniformly 
rotating relativistic polytropes with $n=0.3$ and $0.5$
suggest that the value of $T/|W|$ at the bifurcation point 
strongly depends on compactness $\compa$. In Table \ref{tab:secular}, 
approximate values of quantities 
at the bifurcation point of each model are shown, which are 
evaluated by linearly extrapolating the triaxial sequence to 
the corresponding axisymmetric sequence.  
The value of $T/|W|$ at the bifurcation point becomes  $\sim 0.169$ 
for the compact model $\compa=0.2$, $n=0.3,$ and it will certainly 
 increase for a more compact sequence.  

As seen in the plot of Fig.~\ref{fig:plot_n05}, the triaxial 
solution sequence for $n=0.5$ becomes shorter as  $\compa$
increases.  In fact, we were not able to find a triaxial 
solution sequence for $\compa=0.2$; the triaxial sequence may 
disappear at a certain value of $\compa$ between 0.14-0.2.  
We discuss an interesting consequence of the disappearance of 
triaxial sequences for high compactness in the last section.

\begin{table*}
\begin{tabular}{ccccccccccccc}
\hline
$n$&$\compa$&$R_x$&$R_z/R_x$&$\epsilon_c$&$\Omega$&$\Madm$&$J$&$T/|W|$&$I$&$Z_{\rm p}$ \\
\hline
0.3 & 0.1& $0.3226 $ $(0.3718)$ & $0.5634 $ $(0.5693)$ & $0.3888 $ & $0.6529 $ & $  2.8444\times 10^{-2} $ & $  8.794\times 10^{-4} $ & $0.1507 $ & $  1.347\times 10^{-3} $ & $0.1328 $ & \\
0.3 & 0.14&$0.3435 $ $(0.4220)$ & $0.5531 $ $(0.5619)$ & $0.4445 $ & $0.7161 $ & $  4.4203\times 10^{-2} $ & $  1.903\times 10^{-3} $ & $0.1578 $ & $  2.657\times 10^{-3} $ & $0.2019 $ & \\
0.3 & 0.2& $0.3561 $ $(0.4860)$ & $0.5394 $ $(0.5535)$ & $0.5244 $ & $0.8063 $ & $  6.9865\times 10^{-2} $ & $  4.351\times 10^{-3} $ & $0.1688 $ & $  5.395\times 10^{-3} $ & $0.3311 $ & \\
0.5 & 0.1& $0.5153 $ $(0.5919)$ & $0.5461 $ $(0.5536)$ & $0.2059 $ & $0.4391 $ & $  4.2103\times 10^{-2} $ & $  1.913\times 10^{-3} $ & $0.1493 $ & $  4.356\times 10^{-3} $ & $0.1281 $ & \\
0.5 & 0.12&$0.5209 $ $(0.6169)$ & $0.5439 $ $(0.5531)$ & $0.2319 $ & $0.4698 $ & $  5.2155\times 10^{-2} $ & $  2.753\times 10^{-3} $ & $0.1519 $ & $  5.861\times 10^{-3} $ & $0.1594 $ & \\
0.5 & 0.14&$0.5314 $ $(0.6482)$ & $0.5366 $ $(0.5482)$ & $0.2581 $ & $0.4995 $ & $  6.2304\times 10^{-2} $ & $  3.734\times 10^{-3} $ & $0.1542 $ & $  7.476\times 10^{-3} $ & $0.1932 $ & \\
\hline
\end{tabular}
\caption{Quantities at the point of bifurcation of triaxial sequences from 
axisymmetric sequences.  The polytropic index $n$ and the compactness of 
the spherical star with the same rest mass $\compa$ are the model parameters.  
Corresponding triaxial sequences are found in Table \ref{tab:n03n05all} in 
Appendix \ref{appsec:seq}.  In the above, $R_x$ is the equatorial radius, and 
$R_z/R_x$ is the ratio of polar to the equatorial radius.  Each has two 
values; one is measured in the coordinate length, and the other in 
parenthesis is in proper length defined in Eq.(\ref{eq:proper_length}).  
$\epsilon_c$ is 
the energy density at the center of the compact star, $\Omega$ is 
the angular velocity.  Definitions of $\Madm$, $J$, $T/|W|$, and $I$ 
are found in Appendix \ref{appsec:physq}.  $Z_{\rm p}$ is the polar redshift.
Dimensional quantities are shown 
in $G=c=\kappa=1$ units.}
\label{tab:secular}
\end{table*}

\section{Discussion: proto-neutron star contraction}

As a result of massive stellar core collapses, 
proto-neutron stars are formed and contract to 
more compact neutron stars within the time scale of cooling 
of a few tens of seconds \cite{protoNS}.  
Even for the small rotation rate of the collapsing stellar 
core, the  ratio $T/|W|$ of the proto-neutron star becomes 
much higher than the value where the axisymmetric 
solution becomes secularly unstable against the viscosity driven 
$\ell=m=2$ bar mode instability \cite{LS95,STU04}.  
Therefore, uniformly rotating triaxial solutions 
discussed in this paper may describe a quasi-stationary model of 
proto-neutron star contraction in the range of 
$\compa \sim 0.1-0.2$, assuming the following: 
(1) a certain mechanism of strong viscosity operates during the contraction, 
(2) the time scale rapid cooling is shorter than that of gravitational 
radiation reaction, 
(3) the effective polytropic (adiabatic) index $n$ $(\Gamma=1+1/n)$ of 
the EOS for the realistic neutron star matter is small (large) enough 
to allow uniformly rotating triaxial solutions, 
and 
(4) those triaxial solutions are dynamically stable.  

Such an evolutionary track of a proto-neutron star contraction 
has been considered using a compressible ellipsoidal 
model\footnote{In their work, changes in the entropy during the evolution 
is modeled by the changes in the adiabatic constant $\kappa$ of 
the polytropic EOS.} \cite{STU04}.  
Our results add two further important features to this.  
First, the sequences of triaxial solutions terminate at 
the maximally deformed models, at the mass-shedding limits, 
and the changes in ADM mass or total angular momentum are small 
along the triaxial sequences from the bifurcation points to 
the termination points, even for the stiffer EOS such as $n=0.3$, 
as seen in Table \ref{tab:n03n05all}; the triaxial sequences 
are not very long at all.  
Secondly, the triaxial sequences may become shorter and disappear 
as the compactness becomes larger for a relatively less stiff 
EOS such as $n=0.5$.  

The angular velocity near the braking limit is 
estimated as $\Omega M \sim (M/R)^{3/2} \sim 
(M/R)^{2}J/M^2$.  Therefore once the secular bar mode 
instability sets in, conserving $M$ and $J$, the proto-neutron 
star evolves towards a maximally deformed triaxial 
configuration as it contracts, say, from $M/R\sim 0.1$ to $0.2$.  
And then, it is likely that 
{\it it always evolves along the sequence of maximally 
deformed configurations during the contraction} 
as long as such triaxial solutions exist 
and are dynamically stable in the parameter region of 
the effective $\Gamma$ and $\compa$.  
Excess angular momentum arising from contraction may be 
transported outward by mass ejected from the Lagrange 
point at the cusp of the longest semi-major axis.  (Note that the 
time scale of the mass ejection may be that of cooling, which is 
much longer than the dynamical time scale.)  
Furthermore, if the effective $\Gamma$ satisfies $\Gamma \alt 3$, 
the triaxial solution may disappear when the solution reaches
a certain value of the compactness $\compa$ and higher.  
Dynamically stability of such uniformly rotating solutions 
are not known, but it is unlikely that the dynamical instability 
appears within such short triaxial sequences, along which 
the ratio $T/|W|$ is nearly constant.  

It is estimated that the amplitude of gravitational wave (GW) signals 
from such objects may be detectable using the ground based laser 
interferometric detectors, if the source is within a few tens of 
Mpc \cite{LS95,Saijo06}.  Detection of the persistent GW signals 
even after the proto-neutron star contraction phase suggests a large 
effective $\Gamma \agt 3$, while the shutdown of the signal during 
the contraction implies the relatively smaller $\Gamma \alt 3$.
Detection of such GW signal may set another constraint on the 
EOS parameter of high density matter.  Source modeling for 
constructing the wave templates may be straightforward because 
one can concentrate on calculating the maximally deformed 
configurations.  
Our next plan is to include more realistic nuclear EOS in the code, 
then to estimate the gravitational wave amplitude 
for those EOSs that allow the triaxial solutions.

\acknowledgments
We would like to thank John Friedman for discussions and warm encouragement. 
KU thanks Yoshiharu Eriguchi for discussions and for providing a reprint of 
PhD thesis by Tetsuo Nozawa, and Shin Yoshida for discussions. This work was 
supported by  NSF grants No. PHY0071044, PHY0503366, 
NASA Grant No. NNG05GB99G, the Greek State Scholarships Foundation,  and 
JSPS Grant-in-Aid for Scientific Research(C) 20540275.
%
%
%

\appendix

\section{Formulas for mass and angular momentum}
\label{appsec:physq}

Definitions of the quantities shown in tables and figures that 
characterize each solution of a rotating relativistic star, 
and their expressions in terms of the metric potentials in 
the IWM formulation, are summarized in this Appendix.  

The rest mass of the star $M_0$ is written as 
\beq
M_{0}
\,:=\, \int_\Sigma \rho\,u^\alpha dS_\alpha 
\,=\, \int_\Sigma \rho u^t \alpha \psi^6\sqrt{f}d^3x
\eeq
where $dS_\alpha = \na_\alpha t \sqrt{-g} d^3x$ 
and $\sqrt{-g}d^3x\,=\,\alpha\psi^6\sqrt{f}d^3x$$
\,=\,\alpha\psi^6 r^2 \sin\theta dr d\theta d\phi$.  

The ADM mass $\Madm$ becomes   
\beqn
\Madm
&:=& \frac1{16\pi}\int_\infty 
\left(f^{ac}f^{bd}-f^{ab}f^{cd}\right)\zD_b\gamma_{cd}\, dS_a
\nonumber\\
&=& - \frac1{2\pi}\int_\infty \zD{}^a\psi \,d\zS_a
\,=\, - \frac1{2\pi}\int_\Sigma \zLap\psi \,d\zS
\nonumber\\
&=& \frac1{2\pi}\int_\Sigma \left[\,
\frac18\psi^5\tA_{ab}\tA^{ab}
\,+\,2\pi\psi^5\rhoH \,\right]\sqrt{f}d^3x ,
\nonumber\\
\eeqn
where $d\zS_a = \na_a r \sqrt{f}d^2x$ 
and $d\zS\,=\,\sqrt{f}d^3x$, and $dS_a$ coincides with $d\zS_a$
at spatial infinity.  

The Komar mass associated with a timelike 
Killing field $t^\alpha$ is written   
\beqn
\MK
&:=& -\frac1{4\pi}\int_\infty \na^\alpha\, t^\beta \,dS_{\albe}
\,=\,-\frac1{4\pi}\int_\Sigma \Rab t^\beta\,dS_\alpha
\nonumber\\
&=&  \int_\Sigma \left(\,2\Tab - T \gab \,\right)\,t^\beta \,dS_\alpha, 
\eeqn
and, in the IWM formulation, we have 
\beq
\MK= \int_\Sigma\left[\, \alpha \left(\rhoH+S\right) 
-2 j_a \beta^a\,\right] \psi^6\sqrt{f}d^3x, 
\eeq
where $dS_\alpha = n_\alpha \sqrt{\gamma}d^3x$ was used.  
The above derivation holds if the global timelike 
Killing field exists.  For the spacetime of a triaxially deformed rotating 
star, no such timelike Killing field exists. Instead, an asymptotic Komar 
mass can be written 
\beqn
\MK
&:=& -\frac1{4\pi}\int_\infty \na^\alpha t^\beta \,dS_{\albe}
\,=\,\frac1{4\pi}\int_\infty D^a \alpha\,dS_a
\nonumber\\
&=&  \frac1{4\pi}\int_\Sigma \Dl \alpha\,d\Sigma
\nonumber\\
&=& \frac1{4\pi}\int_\Sigma\left[\, \alpha \tA_{ab}\tA^{ab}
+4\pi\alpha\left(\rhoH+S\right) \,\right] \psi^6\sqrt{f}d^3x .
\nonumber\\
\eeqn
In \cite{SUF04}, we have derived 
sufficient conditions of the fall off of the 
3-metric $\gmabd$ and extrinsic curvature $\Kabd$ 
and their time derivative for the $\Madm = \MK$ relation 
to be satisfied.  In the IWM formulation 
the fall off of each field is sufficiently fast 
to have the equality.  And also in this case, the above 
two definitions for $\MK$ agree.  

The total angular momentum calculated in the asymptotics 
is written
\beqn
J
&:=& -\frac1{8\pi}\int_\infty \pi^a{}_b\phi^b \,dS_a
\,=\,\frac1{8\pi}\int_\infty K^a{}_b\phi^b \,dS_a
\nonumber\\[2mm]
&=&  \frac1{8\pi}\int_\Sigma D_a (K^a{}_b \phi^b)\,dS
\\
&=& \frac1{8\pi}\int_\Sigma 8\pi j_a \phi^a \psi^6\sqrt{f}d^3x.
\eeqn

The relativistic analog of the kinetic energy $T$ is defined by 
\beq
T := \frac12\int\Omega dJ, 
\eeq
therefore, for uniform rotation we have $T = \frac12 \Omega J$.  
Also the relativistic analog of the gravitational potential 
energy $W$ is defined by 
\beq
W := M_{\rm p} + T - \Madm,
\label{eq:grav_pot_en} 
\eeq
where $M_{\rm p}$ is the proper mass defined by 
\beq
M_{\rm p}
\,:=\, \int_\Sigma \epsilon\,u^\alpha dS_\alpha 
\,=\, \int_\Sigma \epsilon u^t \alpha  \psi^6\sqrt{f}d^3x.
\eeq

The proper length of the semi-major axis along the $x$ direction 
is written 
\beqn
\bar{R}_x = \int_0^{R_x} \psi^2 dx, 
\label{eq:proper_length}
\eeqn
where $R_x$ is the coordinate length of the same axis. The proper lengths 
along the $y$ or $z$ directions are calculated using the same
formula, replacing $x$ by $y$ or $z$ respectively.  

In the above, the source terms of the field equations, 
$\rhoH$, $j_a$, and $S$, are obtained from the stress energy tensor.  
We write down the projection of the stress tensor in terms of elementary 
fluid variables and metric potentials.  
The 4-velocity for the corotational flow $u^\alpha = u^t k^\alpha$ 
is decomposed with respect to the foliation $\Sigma_t$ as 
\begin{eqnarray}
u^\alpha n_\alpha &=& -\alpha u^t
\\
u^\alpha \gamma_{\alpha a} &=& u^t \omega_a. 
\end{eqnarray}
Using these relations, the source terms of the field equations 
become 
\beqn
\rhoH \,&:=&\, \Tabd n^\alpha n^\beta 
\,=\, h\rho (\alpha u^t)^2 - p,
\\
j_a \,&:=&\, -\Tabd \gmaa n^\beta
\,=\, h\rho\alpha(u^t)^2 \psi^4\tomega_a,
\\
S \,&:=&\, \Tabd \gamma^{\albe}
\,=\, h\rho\big[(\alpha u^t)^2 - 1\big] +3\,p, 
\eeqn
where $\tomega_a \,:=\, \fabd\omega^b
\,=\, \fabd(\beta^b + \Omega \phi^b) 
\,=\, \tbeta_a + \Omega \tphi_a $.

Throughout the paper we use units such that $G=c=\kappa=1$.
The latter equality is implemented by renormalizing the length and mass scales as
\beq
\bar R\,:=\,\kappa^{-n/2}R,
\qquad
\bar M\,:=\,\kappa^{-n/2}M.
\eeq
respectively. Angular momentum and angular frequency are respectively normalized as
\beq
\bar J\,:=\,\kappa^{-n}J,
\qquad
\bar \Omega\,:=\,\kappa^{n/2}\Omega.
\eeq
We omit the bars over these quantities in the main text.

\section{Selected solution sequences}
\label{appsec:seq}

Sequences of triaxially deformed solutions of the compact star 
were calculated for the following parameters: 
the polytropic index $n=0.3$ with compactness 
$\compa = 0.1, 0.14, 0.2$, 
and 
the polytropic index $n=0.5$ with compactness 
$\compa = 0.1, 0.12, 0.14$. Quantities, most of which are defined in 
Appendix \ref{appsec:physq} 
are tabulated in Table \ref{tab:n03n05all}.  This data is plotted in 
Fig.\ref{fig:plot_n03} and \ref{fig:plot_n05}.

\begin{table*}[!h]
\begin{tabular}{ccccccccccccc}
\hline
\multicolumn{10}{c}{
$ n = 0.3 \quad$  $ \compa = 0.10 \quad$  $ M_0 =  2.9908\times 10^{-2} \quad$  $ M =  2.8116\times 10^{-2} \quad$
}\\
\hline
$R_x$&$R_y/R_x$&$R_z/R_x$&$\epsilon_c$&$\Omega$&$\Madm$&$J$&$T/|W|$&$I$&$Z_p$ \\
\hline
$0.3367 $ $(0.3875)$ & $0.9375 $ $(0.9389)$ & $0.5423 $ $(0.5485)$ & $0.3884 $ & $0.6504 $ & $  2.8446\times 10^{-2} $ & $  8.845\times 10^{-4} $ & $0.1512 $ & $  1.360\times 10^{-3} $ & $0.1327 $  \\ 
$0.3461 $ $(0.3981)$ & $0.8906 $ $(0.8929)$ & $0.5273 $ $(0.5337)$ & $0.3882 $ & $0.6493 $ & $  2.8447\times 10^{-2} $ & $  8.871\times 10^{-4} $ & $0.1515 $ & $  1.366\times 10^{-3} $ & $0.1326 $  \\ 
$0.3567 $ $(0.4098)$ & $0.8438 $ $(0.8470)$ & $0.5115 $ $(0.5181)$ & $0.3880 $ & $0.6475 $ & $  2.8449\times 10^{-2} $ & $  8.909\times 10^{-4} $ & $0.1518 $ & $  1.376\times 10^{-3} $ & $0.1326 $  \\ 
$0.3645 $ $(0.4185)$ & $0.8125 $ $(0.8164)$ & $0.5008 $ $(0.5075)$ & $0.3878 $ & $0.6457 $ & $  2.8450\times 10^{-2} $ & $  8.937\times 10^{-4} $ & $0.1520 $ & $  1.384\times 10^{-3} $ & $0.1325 $  \\ 
$0.3775 $ $(0.4330)$ & $0.7656 $ $(0.7704)$ & $0.4833 $ $(0.4902)$ & $0.3874 $ & $0.6423 $ & $  2.8453\times 10^{-2} $ & $  8.992\times 10^{-4} $ & $0.1524 $ & $  1.400\times 10^{-3} $ & $0.1323 $  \\ 
$0.3875 $ $(0.4441)$ & $0.7344 $ $(0.7397)$ & $0.4707 $ $(0.4778)$ & $0.3871 $ & $0.6397 $ & $  2.8455\times 10^{-2} $ & $  9.043\times 10^{-4} $ & $0.1528 $ & $  1.414\times 10^{-3} $ & $0.1322 $  \\ 
$0.3987 $ $(0.4564)$ & $0.7031 $ $(0.7091)$ & $0.4574 $ $(0.4647)$ & $0.3867 $ & $0.6366 $ & $  2.8458\times 10^{-2} $ & $  9.095\times 10^{-4} $ & $0.1532 $ & $  1.429\times 10^{-3} $ & $0.1320 $  \\ 
$0.4188 $ $(0.4784)$ & $0.6562 $ $(0.6633)$ & $0.4355 $ $(0.4433)$ & $0.3863 $ & $0.6318 $ & $  2.8461\times 10^{-2} $ & $  9.171\times 10^{-4} $ & $0.1536 $ & $  1.452\times 10^{-3} $ & $0.1318 $  \\ 
\hline
\hline
\multicolumn{10}{c}{
$ n = 0.3\quad$   $ \compa = 0.14\quad$  $ M_0 =  4.7471\times 10^{-2}\quad$  $ M =  4.3417\times 10^{-2} \quad$
}\\
\hline
$R_x$&$R_y/R_x$&$R_z/R_x$&$\epsilon_c$&$\Omega$&$\Madm$&$J$&$T/|W|$&$I$&$Z_p$ \\
\hline
$0.3619 $ $(0.4435)$ & $0.9219 $ $(0.9245)$ & $0.5291 $ $(0.5385)$ & $0.4441 $ & $0.7124 $ & $  4.4208\times 10^{-2} $ & $  1.913\times 10^{-3} $ & $0.1581 $ & $  2.686\times 10^{-3} $ & $0.2017 $  \\ 
$0.3763 $ $(0.4603)$ & $0.8594 $ $(0.8639)$ & $0.5089 $ $(0.5187)$ & $0.4439 $ & $0.7103 $ & $  4.4210\times 10^{-2} $ & $  1.919\times 10^{-3} $ & $0.1582 $ & $  2.702\times 10^{-3} $ & $0.2016 $  \\ 
$0.3889 $ $(0.4751)$ & $0.8125 $ $(0.8185)$ & $0.4924 $ $(0.5026)$ & $0.4436 $ & $0.7077 $ & $  4.4213\times 10^{-2} $ & $  1.927\times 10^{-3} $ & $0.1584 $ & $  2.722\times 10^{-3} $ & $0.2014 $  \\ 
$0.3985 $ $(0.4861)$ & $0.7812 $ $(0.7882)$ & $0.4807 $ $(0.4911)$ & $0.4434 $ & $0.7056 $ & $  4.4216\times 10^{-2} $ & $  1.933\times 10^{-3} $ & $0.1586 $ & $  2.739\times 10^{-3} $ & $0.2012 $  \\ 
$0.4091 $ $(0.4984)$ & $0.7500 $ $(0.7579)$ & $0.4684 $ $(0.4792)$ & $0.4431 $ & $0.7032 $ & $  4.4220\times 10^{-2} $ & $  1.940\times 10^{-3} $ & $0.1587 $ & $  2.758\times 10^{-3} $ & $0.2010 $  \\ 
$0.4215 $ $(0.5127)$ & $0.7188 $ $(0.7278)$ & $0.4548 $ $(0.4659)$ & $0.4429 $ & $0.7007 $ & $  4.4223\times 10^{-2} $ & $  1.947\times 10^{-3} $ & $0.1589 $ & $  2.778\times 10^{-3} $ & $0.2008 $  \\
$0.4364 $ $(0.5296)$ & $0.6875 $ $(0.6978)$ & $0.4395 $ $(0.4513)$ & $0.4427 $ & $0.6983 $ & $  4.4226\times 10^{-2} $ & $  1.952\times 10^{-3} $ & $0.1590 $ & $  2.796\times 10^{-3} $ & $0.2007 $  \\
\hline
\hline
\multicolumn{10}{c}{
$ n = 0.3\quad$   $ \compa = 0.20\quad$  $ M_0 =  7.7530\times 10^{-2}\quad$  $ M =  6.7804\times 10^{-2}$
}\\
\hline
$R_x$&$R_y/R_x$&$R_z/R_x$&$\epsilon_c$&$\Omega$&$\Madm$&$J$&$T/|W|$&$I$&$Z_p$ \\
\hline
$0.3745 $ $(0.5091)$ & $0.9219 $ $(0.9262)$ & $0.5162 $ $(0.5313)$ & $0.5242 $ & $0.8036 $ & $  6.9868\times 10^{-2} $ & $  4.360\times 10^{-3} $ & $0.1687 $ & $  5.426\times 10^{-3} $ & $0.3308 $  \\
$0.3819 $ $(0.5184)$ & $0.8906 $ $(0.8967)$ & $0.5064 $ $(0.5219)$ & $0.5242 $ & $0.8026 $ & $  6.9869\times 10^{-2} $ & $  4.362\times 10^{-3} $ & $0.1686 $ & $  5.435\times 10^{-3} $ & $0.3306 $  \\
$0.3945 $ $(0.5343)$ & $0.8438 $ $(0.8524)$ & $0.4905 $ $(0.5066)$ & $0.5241 $ & $0.8008 $ & $  6.9870\times 10^{-2} $ & $  4.369\times 10^{-3} $ & $0.1686 $ & $  5.456\times 10^{-3} $ & $0.3305 $  \\
$0.4227 $ $(0.5691)$ & $0.7656 $ $(0.7791)$ & $0.4584 $ $(0.4762)$ & $0.5240 $ & $0.7968 $ & $  6.9873\times 10^{-2} $ & $  4.380\times 10^{-3} $ & $0.1683 $ & $  5.498\times 10^{-3} $ & $0.3300 $  \\
\hline
\hline
\multicolumn{10}{c}{
$ n = 0.5\quad$   $ \compa = 0.10\quad$  $ M_0 =  4.4113\times 10^{-2}\quad$  $ M =  4.1580\times 10^{-2}$
}\\
\hline
$R_x$&$R_y/R_x$&$R_z/R_x$&$\epsilon_c$&$\Omega$&$\Madm$&$J$&$T/|W|$&$I$&$Z_p$ \\
\hline
$0.5407 $ $(0.6197)$ & $0.9219 $ $(0.9243)$ & $0.5238 $ $(0.5319)$ & $0.2059 $ & $0.4380 $ & $  4.2103\times 10^{-2} $ & $  1.915\times 10^{-3} $ & $0.1492 $ & $  4.373\times 10^{-3} $ & $0.1281 $  \\
$0.5580 $ $(0.6387)$ & $0.8750 $ $(0.8789)$ & $0.5079 $ $(0.5163)$ & $0.2059 $ & $0.4373 $ & $  4.2104\times 10^{-2} $ & $  1.917\times 10^{-3} $ & $0.1491 $ & $  4.384\times 10^{-3} $ & $0.1280 $  \\
$0.5716 $ $(0.6536)$ & $0.8438 $ $(0.8486)$ & $0.4959 $ $(0.5047)$ & $0.2059 $ & $0.4367 $ & $  4.2104\times 10^{-2} $ & $  1.919\times 10^{-3} $ & $0.1491 $ & $  4.393\times 10^{-3} $ & $0.1280 $  \\
$0.5880 $ $(0.6715)$ & $0.8125 $ $(0.8185)$ & $0.4823 $ $(0.4914)$ & $0.2059 $ & $0.4361 $ & $  4.2104\times 10^{-2} $ & $  1.920\times 10^{-3} $ & $0.1490 $ & $  4.402\times 10^{-3} $ & $0.1279 $  \\
\hline
\hline
\multicolumn{10}{c}{
$ n = 0.5\quad$   $ \compa = 0.12\quad$  $ M_0 =  5.5171\times 10^{-2}\quad$  $ M =  5.1345\times 10^{-2}$
}\\
\hline
$R_x$&$R_y/R_x$&$R_z/R_x$&$\epsilon_c$&$\Omega$&$\Madm$&$J$&$T/|W|$&$I$&$Z_p$ \\
\hline
$0.5508 $ $(0.6500)$ & $0.9219 $ $(0.9250)$ & $0.5188 $ $(0.5289)$ & $0.2319 $ & $0.4689 $ & $  5.2155\times 10^{-2} $ & $  2.755\times 10^{-3} $ & $0.1517 $ & $  5.875\times 10^{-3} $ & $0.1594 $  \\
$0.5693 $ $(0.6708)$ & $0.8750 $ $(0.8800)$ & $0.5021 $ $(0.5127)$ & $0.2319 $ & $0.4683 $ & $  5.2156\times 10^{-2} $ & $  2.758\times 10^{-3} $ & $0.1517 $ & $  5.888\times 10^{-3} $ & $0.1594 $  \\
$0.5934 $ $(0.6976)$ & $0.8281 $ $(0.8352)$ & $0.4818 $ $(0.4932)$ & $0.2319 $ & $0.4677 $ & $  5.2156\times 10^{-2} $ & $  2.759\times 10^{-3} $ & $0.1516 $ & $  5.900\times 10^{-3} $ & $0.1593 $  \\
\hline
\hline
\multicolumn{10}{c}{
$ n = 0.5\quad$   $ \compa = 0.14\quad$  $ M_0 =  6.6547\times 10^{-2}\quad$  $ M =  6.1130\times 10^{-2}$
}\\
\hline
$R_x$&$R_y/R_x$&$R_z/R_x$&$\epsilon_c$&$\Omega$&$\Madm$&$J$&$T/|W|$&$I$&$Z_p$ \\
\hline
$0.5410 $ $(0.6591)$ & $0.9688 $ $(0.9703)$ & $0.5282 $ $(0.5401)$ & $0.2580 $ & $0.4993 $ & $  6.2304\times 10^{-2} $ & $  3.735\times 10^{-3} $ & $0.1542 $ & $  7.482\times 10^{-3} $ & $0.1932 $ & \\
$0.5630 $ $(0.6843)$ & $0.9062 $ $(0.9109)$ & $0.5077 $ $(0.5203)$ & $0.2580 $ & $0.4988 $ & $  6.2305\times 10^{-2} $ & $  3.738\times 10^{-3} $ & $0.1542 $ & $  7.495\times 10^{-3} $ & $0.1931 $ & \\
$0.5771 $ $(0.7003)$ & $0.8750 $ $(0.8813)$ & $0.4954 $ $(0.5085)$ & $0.2580 $ & $0.4984 $ & $  6.2306\times 10^{-2} $ & $  3.740\times 10^{-3} $ & $0.1542 $ & $  7.505\times 10^{-3} $ & $0.1931 $ & \\
\hline
\end{tabular}
\caption{
Quantities for the constant rest mass Jacobi-like triaxial sequences.  
$M_0$ is the rest mass of each sequence, and $M$ is the gravitational 
mass of the spherical star having the same $M_0$.  In the definition 
of the compactness, $R$ is the Schwartzschild radius of the spherical star.  
Tabulated quantities are the same as Table \ref{tab:secular}, except for 
$R_x$, which is the semi-major radius along the $x$-axis, and $R_y$, 
the semi-major radius along the $x$-axis.  
Dimensional quantities are shown in $G=c=\kappa=1$ units.}
\label{tab:n03n05all}
\end{table*}

\end{document}